\documentclass{article}

\usepackage{amssymb}
\usepackage{booktabs}
\usepackage{graphicx}
\usepackage{lipsum}
\usepackage{caption}
\usepackage{subcaption}
\usepackage{amsmath}
\usepackage{multirow}
\usepackage{longtable}
\usepackage{makecell}
\usepackage{tabularx}
\usepackage{floatrow}
\usepackage{xltabular}
\usepackage{color}
\usepackage{tabularray}
\usepackage{threeparttable}
\usepackage{tablefootnote}
\usepackage{tcolorbox}
\usepackage{upgreek}
\usepackage{acronym}
\usepackage{hyperref}
\usepackage{enumitem}
\usepackage{pdfpages}
\usepackage{lineno}
\usepackage{comment}
\usepackage[usestackEOL]{stackengine}
\usepackage{algorithm}
\usepackage{algorithmicx}
\usepackage[noend]{algpseudocode}
\usepackage{appendix}
\usepackage{chngcntr}
\usepackage{etoolbox}

\usepackage{amsmath}  
\usepackage{textcomp} 
\usepackage{siunitx} 

\usepackage{geometry}

\graphicspath{{Figures/}}
\hypersetup{
    colorlinks = true,
    linkcolor = blue,
    urlcolor  = blue,
    citecolor = blue,
    anchorcolor = blue
}

\usepackage{float}
\usepackage{here}
\usepackage{acronym}

\usepackage[sorting=none, style=numeric-comp]{biblatex}

\usepackage{authblk}

\title{Extension of a phase-field KKS model to predict the microstructure evolution in LPBF AlSi10Mg alloy submitted to non isothermal processes}

\author[a]{Seifallah Fetni}
\author[b]{Jocelyn Delahaye}
\author[a]{Héctor Sepúlveda}
\author[a]{Laurent Duchêne}
\author[a]{Anne Marie Habraken}
\author[b]{Anne Mertens\thanks{Corresponding author e-mail: anne.mertens@uliege.be}}

\affil[a]{University of Liège, UEE Research Unit, MSM division, allée de la Découverte, 9 B52/3, B 4000 Liège, Belgium}
\affil[b]{University of Liège, Aerospace \& Mechanics, MMS Unit, allée de la Découverte, 9 B52/3, B 4000 Liège, Belgium}

\date{}
\begin{document}
\maketitle


\begin{abstract}
The out-of-equilibrium heterogeneous microstructure typical of AlSi10Mg processed by Laser Powder Bed Fusion (LPBF) is often modified by further heat treatment to improve its ductility. According to literature, extensive experimental investigations are generally required in order to optimize these heat treatments. In the present work, a phase-field approach is developed based on an extended Kim-Kim-Suzuki (KKS) model to guide and accelerate the post-treatment optimization. Combined with CALculation of PHAse Diagrams (CALPHAD) data, this extended KKS model predicts microstructural changes under anisothermal conditions. To ensure a more physical approach, it takes into account the enhanced diffusion by quenched-in excess vacancies as well as the elastic energy due to matrix/precipitate lattice mismatch. As the developed model includes the computation of the evolution of the thermo-physical properties, its results are validated through comparison with experimental DSC curves measured during the non-isothermal loading of as-built LPBF AlSi10Mg. The computed microstructure evolution reproduces the microstructural observation and successfully explains the peaks in the DSC heat flow curve. It thus elucidates the detailed microstructural evolution inside the eutectic silicon phase by considering the growth and coalescence of silicon precipitates and the matrix desaturation.
\end{abstract}
\textit{\textbf{Keywords}}:
KKS, LPBF, Phase-Field modeling, CALPHAD, AlSi10Mg

\section{Introduction}
\par In Laser Powder Bed Fusion (LPBF) process, thermal cycles and solidification velocities are considerably increased compared to classical manufacturing process and ordinary directional solidification. The result is an out-of-equilibrium microstructure mainly characterized by very fine cellular, dendritic and inhomogeneous phases. During the LPBF process, various phenomena such as heat transfer, fluid flow, moving boundaries and crystalline anisotropy occur \cite{DELAHAYE2019160,KIMURA20161294}. The built platform is often heated before the process starts and a noticeable heat accumulation in the built part, as result of the laser power, occurs during the process. This thermal energy triggers the transformation of the generated out-of-equilibrium microstructure towards an equilibrium state. The resulting complex material thermal history and microstructure evolution explain why  advanced numerical methods are required to predict the thermo-physical properties during and after the LPBF process.

\par Among the LPBF powders, AlSi10Mg alloy has become extremely attractive thanks to its good weldability, low shrinkage as well as high fluidity. However, the process development still faces various challenges, especially the high anisotropy in strength and the lack of ductility of the manufactured parts \cite{BAO2022108215}. Experimental investigations have demonstrated the strong influence of the molar fraction, morphology and size of Si particles on the mechanical properties of the parts \cite{Hegde2008} (as-built and/or heat treated). Indeed, circular and coarse Si particles can act as crack initiation sites resulting in low ductility. Therefore, the refinement of the eutectic Al-Si microstructure has been a main target in several works \cite{LI201574,LIAO20071121,SHANKAR20044447}. 
However, most of the studies are experimental ones, which means a significant delay and cost. For instance, Yang et al. \cite{YANG202278} have recently applied DSC measurements to unravel precipitation phenomena in Al-Mg-Si alloys, putting the emphasis on the intricacies of non-isothermal heating and highlighting the crucial role of excess vacancies in the early stages of the process. The implementation and development of numerical tools capable of reliable predictions would be of great interest to deal with alloys of the Al-Mg-Si system.

\par Phase-field modeling is a powerful tool to simulate microstructure evolutions in metallurgy. It assumes that the phases are separated by smooth interfaces with finite widths (typically few nanometres). This interface model avoids handling complex boundary conditions at the moving interfaces. The method considers the interfacial energy between phases, as a key physical parameter required for the model parametrization. By minimizing the total free energy, the evolution of the concentration and phase-fields are determined as well as the physical interface width and the interface energy.
The phase-field method has been developed and extended to be one of the most promising approaches applied in computational materials science. It can deal with microstructural changes in materials under high thermal gradients, and is now integrated into materials engineering for improved material design \cite{LASKOWSKI2021158630}. Such method has allowed the study of various complex transformations under severe conditions. One can enumerate solidification \cite{GU2021110812,LINDROOS2022103139}, coarsening of precipitates \cite{BOISSE20076151}, ferromagnetic solid materials \cite{ZHANG2020113310}, phase changes \cite{HAGHANIHASSANABADI2021110111}, spinodal decomposition \cite{FETNI2023111820} and crack propagation \cite{SCHOLLER2022114965}.
\par Initially developed to compute phase equilibrium in multicomponent and multiphase systems, CALculation of PHAse Diagrams (CALPHAD) evaluates the Gibbs free energy of single phases by simple polynomial functions \cite{Campbell2014}. The thermodynamic equilibrium of a material is calculated by minimizing its Gibbs free energy \cite{YANG2023111}. The method has been further extended to model properties such as diffusion mobility of the chemical species, elastic constants, molar volumes... and more recently the density and thermal conductivity of a system. For instance, based on the microstructure of AL-Si alloys, a Calphad model is described to compute the conductivity depending on temperature \cite{Zhang2016}.
CALPHAD thermodynamic simulations have been applied on additively manufactured alloys to predict for instance equilibrium phase contents \cite{KATZDEMYANETZ2020110505}, microsegregation and solidification microstructure \cite{KELLER2017244}. Once the equilibrium state is known, the phase fractions, their compositions as well as other thermodynamic quantities are derived from the Gibbs free energy.
\par The so-called Kim-Kim-Suzuki (KKS) model, developed to deal with the solidification of binary alloys, was introduced in \cite{KKS1999}. The interface is treated as a mixture of the two phases with equal chemical potentials which means that the interfacial energy is decoupled from the interface width. As a consequence, the interface width between the two phases can be artificially increased, extending the simulation length scale and allowing running simulations in a “reasonable” computing time and memory capacity. Such a model is quite interesting to provide quantitative predictions from the phase-field approach. Furthermore, the KKS model relies on physical-based parameters which are gathered from experiments or calculated. It has been extended to the solid state transformations by including the elastic strain effect. For instance, Hu et al. \cite{HU2007303} investigate the effect of different approximations of chemical free energies to predict the growth and kinetics of plate-like $\theta’$ (Al$_2$Cu) in Al–4wt\%Cu alloys, while Ji et al. \cite{JI201884} focus on the morphology of those $\theta’$ precipitates in 319 Al-alloy (Al-3.5wt\%Cu-6.0wt\%Si).

\par
\begin{figure*}[htbp]
    \centering
    \includegraphics[width=0.85\textwidth]{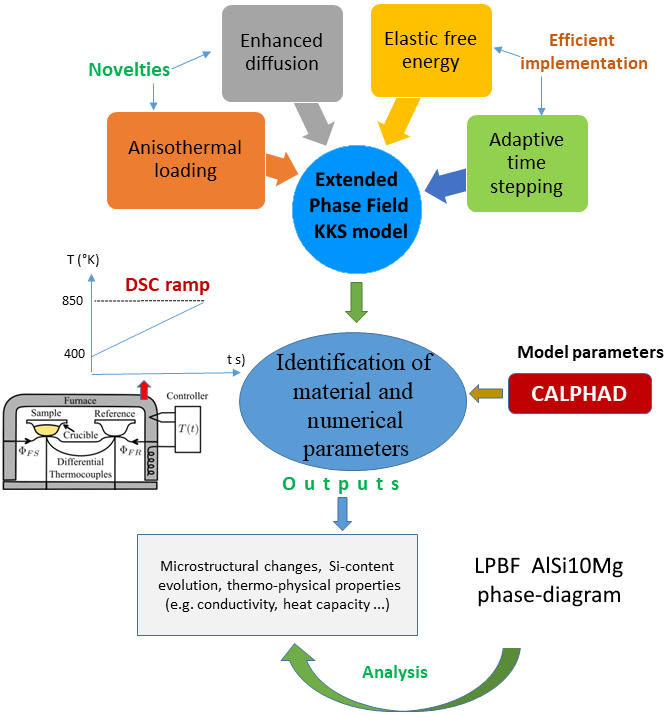}
    \caption{Flowchart summarizing the methodology of the present work and enumeration of the predicted outputs.}
    \label{fig:flowchart}
\end{figure*}

\begin{figure}[htbp]
    \centering
    \includegraphics[width=0.5\textwidth]{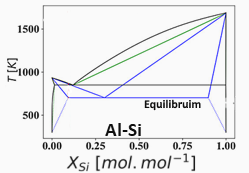}
          \caption{Phase diagram of the Al-Si binary system.  Al-Si phase diagram (black line), Al-Si kinetic phase diagram for an
interface velocity of 0 m.s$^{-1}$ (green) and 0.2 m.s$^{-1}$ (blue). } 
\label{fig:AlSi}
\end{figure}

 In the present research, an enhanced KKS model is developed to predict the growth and the coarsening kinetics of Si precipitates in AlSi10Mg LPBF alloy during a heat ramp. This novel KKS implementation takes into account the elastic free energy associated to matrix/precipitate mismatch as well as an enhanced diffusion by quenched-in excess vacancies, and  it uses an adaptive time stepping approach. It can follow an anisothermal load (in this case a 20K/min ramp). The methodology of the present work and its outputs are summarized in Fig. \ref{fig:flowchart} and the phase diagram of the Al-Si binary system is illustrated in  Fig. \ref{fig:AlSi}. Note that the phase-field equations are detailed hereafter with the origin of the input data. They consist in the total free energy, the atomic mobility and inter-diffusivity, the molar volume, the eigen strain, the stiffness tensor, the interfacial energy and the enhanced diffusion by quenched-in excess vacancies. Applied to simulate a Differential Scanning Calorimetry experiment, the model is validated by comparing its outcomes with experimental ones. It quantifies the microstructural changes of the as-built LPBF AlSi10Mg post processed by a thermal treatment. Many material features are predicted such as the molar fraction of the alloying elements, morphology and size of the diamond Si phases, thus paving the way towards the accurate prediction of thermophysical properties, both during or after the LPBF process. 

\section{Sample manufacturing, post treatment and characterization} \label{exp_section}
\par The chemical composition of the LPBF AlSi10Mg, measured by Inductively Coupled Plasma Optical Emission Spectroscopy (ICP-OES), is as follows: Si: 9.68 wt\%, Mg: 0.43 wt\%, and Al: Balance (Bal).
For Differential Scanning Calorimetry (DSC) measurements, vertical cylinders of 19 mm in diameter and 100 mm in height were built with an EOS M290 LPBF machine with the following parameters: laser power of 370 W, laser scan speed of 1.3 m/s, powder layer thickness of 30 $\mu$m, hatch spacing of 0.19 mm, 67$^{\circ}$ rotation between the layers and a built platform temperature of 35$^{\circ}$C. Cylinders were produced in one batch of 37 and the fabrication time reaches 54 hours for the batch. DSC specimens were extracted from these cylinders as disks with a diameter of 5 mm and a thickness of 1 mm.
\par
\begin{figure*}[htbp]
\centering
\subcaptionbox{ \label{before_DSC}}
{\includegraphics[width=0.45\textwidth]{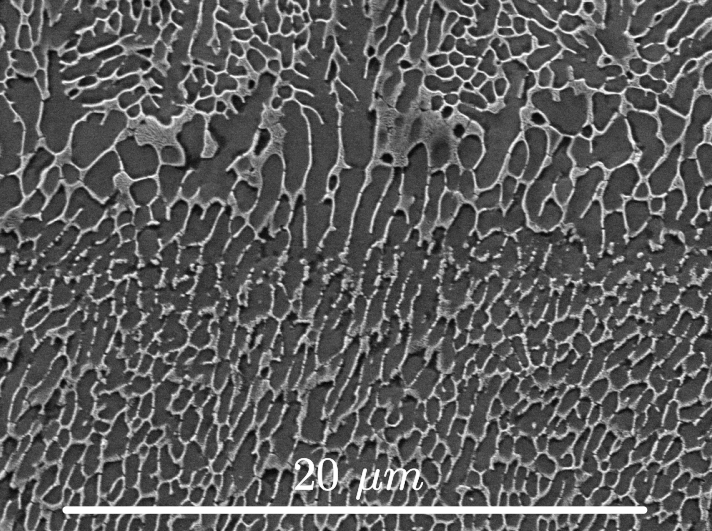}}
\hspace{0.5cm}
\subcaptionbox{ \label{after_DSC}}
{\includegraphics[width=0.45\textwidth]{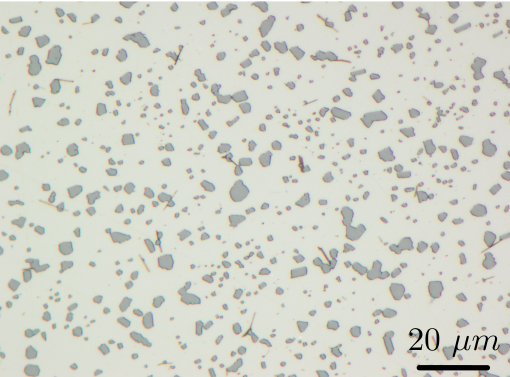}}

\subcaptionbox{ \label{liquid_state}}
{\includegraphics[width=0.45\textwidth]{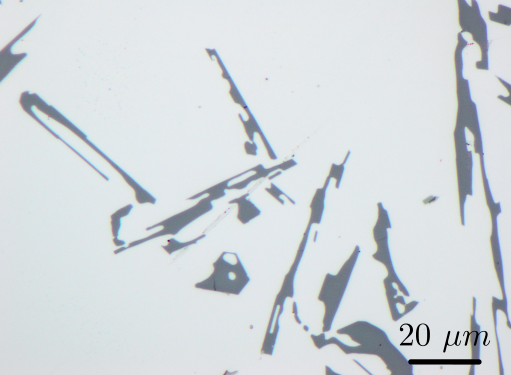}}
\caption{Microstructure of an as-built AlSi10Mg LPBF characterized by optical microscope, and
SEM characterization of (b) an annealed AlSi10Mg LPBF microstructure with globularized Si
precipitates and (c) an as cast microstructure after a slow cooling from the melt.}
\label{fig:SEM_micrographs}
\end{figure*}

 Samples in as-built condition and after DSC tests were observed after polishing and etching with Keller reagent (95\% distilled water, 2.5 \% HNO$_{3}$, 1.5 \% HCl, 1 \% HF, \% in volume). The microstructure of as-built AlSi10Mg LPBF is composed of different zones, related to their location in the melt pool (see Fig.\ref{fig:SEM_micrographs}). They are called the Melt Pool fine (MP fine) within the melt pool core, the Melt Pool coarse (MP coarse) close to the melt pool boundary and the Heat Affected Zone or HAZ where the microstructure is modified by the heat brought when the subsequent layers are printed. Hereafter, only MP fine zone is studied and modeled with the KKS model as it is the main component of the microstructure as characterized in \cite{DELAHAYE2019160}. This MP fine microstructure has an average cell size of 360 nm and consists in an Al-Si supersaturated solid solution ($X_{Si}^{\alpha_{0} }$=0.025 initial Si concentration within Al-Si solid solution called $\alpha$ phase) with a weight fraction of dispersed diamond Si phase precipitates of 0.075. These $\alpha$-Al cells are separated by eutectic mixtures containing Si precipitates with an average size of 20 nm, as characterized by in-house XRD measurements.
\par The microstructures of the AlSi10Mg LBPF in as-built state, after an annealing heat treatment and after the DSC test at the liquid state are shown in Figure \ref{fig:SEM_micrographs}[(a), (b) and (c)] respectively. One can see that the as-built microstructure exhibits sub-micron $\alpha$-Al primary cells surrounded by an Al-Si eutectic mixture. After heat treatment, the Si precipitates have grown and coarsened into globular precipitates of several microns in diameter. When the AlSi10Mg LPBF in as-built state is heated up to the liquid state and then cooled down in the DSC apparatus, a typical as-cast microstructure is obtained with plate-like Si precipitates of ten of microns in length in the eutectic mixture and a few primary $\alpha$ dendrites. Note the scale difference between the as-built microstructure and the ones obtained after heat treatments.
\par
\begin{figure*}[htbp]
\centering
\subcaptionbox{ \label{}}
{\includegraphics[width=0.45\textwidth]{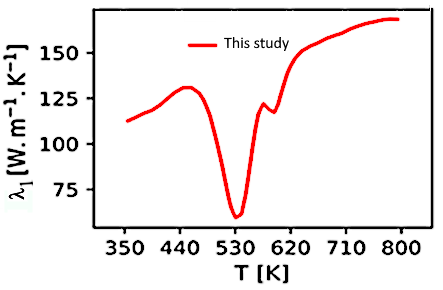}}
\subcaptionbox{ \label{}}
{\includegraphics[width=0.45\textwidth]{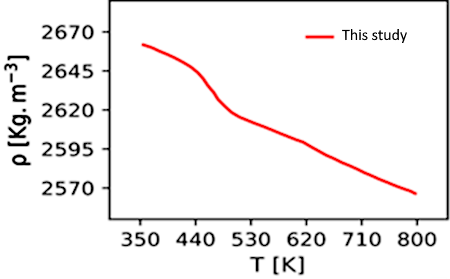}}

\subcaptionbox{ \label{}}
{\includegraphics[width=0.45\textwidth]{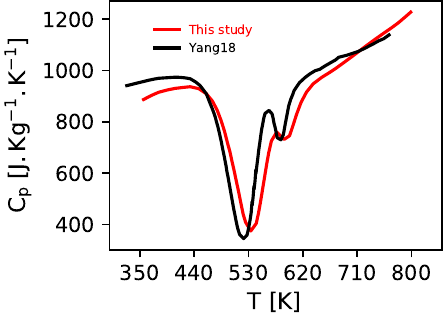}}
\caption{Measured thermo-physical properties for AlSi10Mg LPBF : (a) Thermal conductivity $\lambda_{1}$ , (b) density $\rho$ and (c): heat capacity C$_{p}$ compared with the results of \cite{yang_deibler_bradley_stefan_carroll_2018} .}          
\label{fig:properties}
\end{figure*}

 Sample density as a function of temperature was calculated based on the coefficient of linear expansion measured by a Netzsch DIL 402 C dilatometer with a heating rate of 3 K/min and the density measured at ambient temperature by a Micromeritics AccuPyc II 1340 pycnometer. 
The heat capacity, the fusion enthalpy, the solidus and liquidus temperatures were determined by a Netzsch DSC 404 C Pegasus Differential Scanning Calorimetry with a heating rate of 20 K/min under Argon atmosphere.
The thermal diffusivity was measured by a Netzsch LFA 427 laser flash apparatus under Argon atmosphere. The test was performed with a heating rate of 5 K/min. The thermal conductivity $\lambda_{1}(T)$ was computed from the thermal diffusivity $\alpha_{1}$, the density $\rho$ and the heat capacity C$_{p}$ according to the formula:
\begin{equation}
\lambda_{1}(T)=\alpha_{1}(T)~\rho(T)~C_{p}(T)
\label{lambda_rou_alpha_}
\end{equation}
where \textit{T} is the temperarture. 
The measurement results are presented in Fig. \ref{fig:properties}, and for the thermal capacity, compared with literature values. 
\section{Extended Phase Field KKS model} \label{phase_field_model}

In the applied phase-field approach, for the bi-phasic material state associated to the as-built or the post treated AlSi10Mg LPBF, the conservative field is the total quantity of Al and Si components, while the non-conservative field also called order parameter defines the phases: either the Al-Si solid solution or the pure diamond Si precipitates dispersed inside $\alpha$-Al cells or present within eutectic wall. In section \ref{free_enrg}, the equations used to compute the system energy are detailed, while section \ref{kinet_eq} describes the two main equations (Cahn-Hilliard (CH) and Allen-Cahn (AC)) allowing the computation of the system evolution.
The KKS model was developed for simulating binary alloy solidification under isothermal conditions \cite{KKS1999}. However, in the present work, several significant extensions have been made to the initial KKS model to enable its application to the non-isothermal microstructure evolution in LPBF-processed AlSi10Mg alloys.
One of the key novelties is the incorporation of temperature dependence in the bulk free energy expression (Eq. \ref{eq:eq2}), where $f_0$ is now a function of the order parameter $\eta$, the Si molar fraction $X_{\text{Si}}$, and the temperature $T$. This temperature dependence is not just qualitative but also quantitative, as it is integrated into the total free energy formulation $F_{\text{KKS}}$ (Eq. \ref{eq:eq1}) and subsequent equations (e.g., Eq. \ref{eq:res_kks}). This extension allows the model to handle anisothermal conditions, such as those encountered during heat treatments or characterization techniques like differential DSC.
Another important extension is the consideration of the enhanced diffusion due to quenched-in excess vacancies generated by the rapid cooling rates during the LPBF process. This effect is incorporated through a modified diffusion coefficient expression (Eq. \ref{e03_diffu_effec}), which includes a term for the fraction of excess vacancies $X_{\text{Va}}$. By accounting for this enhanced diffusion, the model can better capture the microstructural evolution kinetics in the LPBF-processed material.
Furthermore, the implementation of the KKS model in this work employs an adaptive time-stepping approach. This adaptive time-stepping strategy aims to improve the numerical stability and efficiency of the simulations, particularly when dealing with the complex microstructural changes occurring during non-isothermal processes.
To further enhance numerical efficiency and stability, we derive dimensionless forms of the CH and AC equations (Eqs. \ref{e01_cahn_norma} and \ref{e02_allen_norma} , respectively). This dimensionless formulation helps in ensuring efficient computing and facilitates the numerical implementation of the model.
\subsection{Key assumptions for the extended Phase Field KKS model} 
While the Al-Si binary phase diagram in Fig. \ref{fig:flowchart} does not fully represent the ternary Al-Si-Mg system, the relatively small Mg content (0.43 wt\%) in AlSi10Mg alloy and its potential loss during the high-temperature LPBF process suggest a limited impact on phase equilibrium and microstructure evolution compared to the dominant Al-Si system. Furthermore, the absence of Mg-containing precipitates in the as-built LPBF microstructure indicates that Mg likely remains in solid solution within the Al matrix. Considering the original KKS model's formulation for binary alloys and the current work's focus on its fundamental aspects and extension to handle anisothermal conditions, using a binary Al-Si model is a reasonable simplification. This approach aligns with the common practice of neglecting the influence of a third element on phase formations when its impact is expected to be negligible within the given conditions, as demonstrated in the study by 
Keller et al. \cite{KELLER2017244} on Nb segregation in laser additive manufacturing of the multicomponent Inconel 625 nickel-based superalloy, where a simplified Ni-Nb binary system was used for phase-field simulations.
\subsection{Free energy} \label{free_enrg}
In the new KKS model, the Helmholtz free energy $\mathcal{F}_{KKS}$ of the system taking into consideration the elastic misfit strain is expressed as: 
\begin{equation}
\begin{aligned}
\mathcal{F}_{KKS}(X_{Si},\eta,T)= \langle  V_{m} \rangle 
\int_{V}\left[
f_{0}(X_{Si},\eta,T)+\dfrac{\kappa^{2}}{2} (\nabla \eta)^{2} + e_{el}(\eta) \ \right] dV
\label{eq:eq1}
\end{aligned}
\end{equation}
where $\left\langle V_{m}\right\rangle$ is the mean molar volume of the system, $X_{Si}$ the molar fraction of Si in the system, $f_{0}$ the bulk free energy density, $\kappa$ the gradient energy coefficient, $e_{e l}$ the elastic energy due to the mismatch between the matrix and the precipitate. $\eta$ is the order parameter which represents the d-Si diamond phase. Therefore,  d-Si precipitate is present if $\eta$ = 1
and the $\alpha$-Al faced-centred cubic (FCC) Si-Al solid solution is present if $\eta$ = 0.	
\newline The bulk free energy $f_{0}$ i is expressed as the molar free energy of the single
phases to which is added a potential at the interface as follows:	
\begin{equation}
\begin{aligned}
f_{0}(X_{Si},\eta,T)=
\dfrac{1}{\langle V_{m} \rangle }
\left[ (1-h(\eta))~f^{\alpha}(X_{Si},T)+h(\eta)~f^{d}(X_{Si},T)+wg(\eta) \right]
\label{eq:eq2}
\end{aligned}
\end{equation}
where $f^{\alpha}$ and $f^{d}$ are the molar free energy of the $\alpha$-Al matrix and the d-Si precipitate
phases respectively and \textit{w} is the potential height for the phase interface. \textit{h} represents an interpolation function and \textit{g} a double-well potential function. These functions are expressed as:
	\begin{equation}
		\begin{aligned}
h(\eta)=3 \eta^{2}-2 \eta^{3}      
		\label{eq:eq3}
				\end{aligned}
		\end{equation}
	\begin{equation}
			\begin{aligned}
g(\eta)= \eta^{2}-2 \eta^{3} +\eta^{4} 
		\label{eq:eq4}	
			\end{aligned}
	\end{equation}
The interpolation function $h(\eta)$, already used in Eq. \ref{eq:eq2} to compute the system energy, is also exploited to define the concentration of Si in the system as an interpolation of the order parameter as:  
	\begin{equation}
			\begin{aligned}	
X_{S i}=(1-h(\eta)) X_{S i}^{\alpha}+h(\eta) X_{S i}^{d}
		\label{eq:eq5}	
			\end{aligned}		
	\end{equation}
where $X_{S i}^{\alpha}$ and $X_{S i}^{d}$ represent the concentration of Si in $\alpha$ and diamond phases respectively. These phases have different compositions but the same diffusion potential:
	\begin{equation}
	\frac{\partial f^{\alpha}}{\partial X_{Si}^{\alpha}}=\frac{\partial f^{d}}{\partial X_{Si}^{d}}
	\label{eq:eq6}	
	\end{equation}
It follows that the gradient energy coefficient $\kappa$ and the double-well potential height \textit{w}
can be expressed as (\cite{HU2007303}):
	\begin{equation}
			\begin{aligned}	
\kappa^{2}=\frac{3}{\alpha} \gamma 2 \lambda 
			\end{aligned}		
				\label{eq:eq7}	
	\end{equation}
	\begin{equation}
			\begin{aligned}	
w=3 \frac{\gamma \alpha}{\lambda}
			\end{aligned}		
				\label{eq:eq8}	
	\end{equation}
where $\gamma$ is the interface energy, $\alpha$ is a constant determined by the interface definition, whose value is equal to 2.2 and 2$\lambda$ is the interface thickness.
\newline The elastic energy arising from the coherency or misfit strains between the matrix and the precipitate reads :
	\begin{equation}
			\begin{aligned}	
e_{e l}=\frac{1}{2} C_{i j k l}\left(\varepsilon_{i j}-\varepsilon_{i j}^{0}\right)\left(\varepsilon_{k l}-\varepsilon_{k l}^{0}\right)
			\end{aligned}		
				\label{eq:eq9}	
	\end{equation}
where $C_{i j k l}$ is the stiffness tensor, $\varepsilon_{i j}^{0}$ the eigen strain or stress-free strain, $\varepsilon_{i j}$ the total strain. The total strain is solved by the stress equilibrium equation:
	\begin{equation}
\left\{\begin{aligned}
\nabla_{j}~ \sigma_{i j} &=0 \\
\sigma_{i j} ~~&=C_{i j k l}\left(\varepsilon_{k l}-\varepsilon_{k l}^{0}\right)
\end{aligned}\right.	
				\label{eq:eq10}	
	\end{equation}
\subsection{Kinetic equations }  \label{kinet_eq}
On the one hand, the temporal evolution of the conserved field $X_{S i}$ is governed by CH equation: 
\begin{equation}
\begin{aligned}
\dfrac{\partial X_{Si}}{\partial t}
=& \nabla.
\left (
M~\nabla \dfrac{\delta \mathcal{F}_{KKS}}{\delta X_{Si}}
\right)
\label{e01_cahn_hilli}
\end{aligned}
\end{equation}
where \textit{t} is the time, \textit{M} the diffusion mobility between $\alpha$-Al and d-Si phases. This parameter is computed by Eq. \ref{e03_mobil}. 
\begin{equation}
\begin{aligned}
M=& \langle V_{m} \rangle^{2}~X_{Al}^{\alpha}~X_{Si}^{\alpha}~(X_{Al}^{\alpha}~M_{Si}^{\alpha}
+X_{Si}^{\alpha}~M_{Al}^{\alpha}  
)  \\
=&\dfrac{\tilde{D}}{\partial^{2} f_{0} / \partial X_{Si}^{2}}
\label{e03_mobil}
\end{aligned}
\end{equation}
where $\tilde{D}$ is the inter-diffusivity.  
\newline On the other hand, the evolution of the non-conserved field $\eta$ is governed by AC equation :
\begin{equation}
\begin{aligned}
\dfrac{\partial \eta}{\partial t}
=& -L\dfrac{\delta \mathcal{F}_{KKS}}{\delta \eta}
\label{e02_allen_cahn}
\end{aligned}
\end{equation}
where \textit{L} is the interface kinetic coefficient. It can be derived through the so-called thin-interface analysis from \cite{KKS1999}.
\begin{equation}
\begin{aligned}
L
=& 
\dfrac{\gamma}{\kappa^{2}}
\dfrac{1}
{
\frac{1}{M_{\eta}}+\frac{\kappa \xi}{\tilde{D}\sqrt{2w}}    
}
\label{e04_kinet_coeff}
\end{aligned}
\end{equation}
with:
\begin{equation}
\begin{aligned}
\xi
=& 
\dfrac{1}{\langle V_{m} \rangle}~
f^{\alpha}_{X_{Si}X_{Si}}(X_{Si}^{\alpha_{e}})~
f^{d}_{X_{Si}X_{Si}}(X_{Si}^{d_{e}})~
(X_{Al}^{\alpha_{e}}-X_{Al}^{d_{e}})^{2}  \\
&\times \int_{\eta}^{}
\dfrac{h(\eta)~(1-h(\eta))}
{
(1-h(\eta))~f^{\alpha}_{X_{Si}X_{Si}}(X_{Si}^{\alpha_{e}})
+h(\eta)~f^{d}_{X_{Si}X_{Si}}(X_{Si}^{d_{e}})
} \,
 \dfrac{1}{\eta(1-\eta)}d\eta
\end{aligned}
\label{e05_thin_inter_analy}
\end{equation}

In equations \ref{e04_kinet_coeff} and \ref{e05_thin_inter_analy} , $M_{\eta}$ is the interfacial mobility and $f_{X_{S i} X_{S i}}^{\alpha}$ is the short-hand notation for $\frac{\partial^{2} f^{\alpha}}{\partial X_{S i}^{2}}$. The \textit{e} within $\alpha_{e}$ or $d_{e}$ notations expresses the equilibrium state. 
Note that, equation \ref{e05_thin_inter_analy} can be simplified by assuming that $f_{X_{S i} X_{S i}}^{\alpha}=f_{X_{S i} X_{S i}}^{d}$   and becomes:
\begin{equation}
\begin{aligned}
\xi
=& 
\dfrac{1}{\langle V_{m} \rangle}
f^{\alpha}_{X_{Si}X_{Si}}(X_{Si}^{\alpha_{e}})
(X_{Si}^{\alpha_{e}}-X_{Si}^{d_{e}})^{2} 
\dfrac{19}{30}
\label{e06_thin_inter_simpl}
\end{aligned}
\end{equation}
It is worth noting that, while the Gibbs free energy curve width for the d-Si phase is indeed narrower than that of the $\alpha$-Al phase, the d-Si phase can be considered as a nearly stoichiometric compound with a very low solubility range for Al. As a result, the second derivative of the free energy with respect to the Si molar fraction ($f_{X_{Si}X_{Si}}^\alpha$) may have a negligible influence on the driving forces for precipitation and coarsening processes within the d-Si phase. This argument finds support in the work of Hu et al. \cite{HU2007303} and the associated Thesis \cite{hu2004phase}, where they performed an analysis considering the differences in free energy curve widths between the matrix and precipitate phases, by considering $\theta$ in Al-Cu alloys. 
\section{Material parameters and CALPHAD data}   \label{param_calphad}
Hereafter, the approach to identify each input material parameter of the phase-field model, defined in section \ref{phase_field_model}, is presented. 
\subsection{Free energy } 
In solid-state transformations, the molar volume variation is small, and the work pressure force can be neglected. As a result, the molar Gibbs free energy \( G_{m}^{\phi} \) of a phase \( \phi \) is a good approximation of its Helmholtz free energy \( f^{\phi} \), such that \( f^{\phi} \approx G_{m}^{\phi} \).

The molar Gibbs free energies of $\alpha$-Al and d-Si ($\phi$= $\alpha$, d-Si) are modeled by the Redlich-Kister formalism in the COST 507 open database for light metal alloys developed in \cite{ADR98}.
To lighten the computation in the current work, parabola functions were fitted as approximation of the Gibbs free energies. Indeed, if one directly uses the Redlich-Kister formalism in the phase-field model, Eq. \ref{eq:eq6} becomes non-linear
and has to be solved numerically, which can be time consuming for large systems. Hu et al. \cite{HU2007303} have shown that a good approximation of Gibbs free energy is obtained by fitting a parabola ($^P G_m^\phi$) minimizing the transformation driving force for $X_{Si}^{\alpha_{e}}$ $<$ $X_{Si}^{\alpha_{}}$ $<$ $X_{Si}^{\alpha_{0}}$ at a given temperature. $X_{Si}^{\alpha_{e}}$ and $X_{Si}^{\alpha_{0}}$ are the molar fraction of Si in $\alpha$-Al at equilibrium and for the initial state $\alpha_{0}$. The parabola function reads:
\begin{equation}
\begin{aligned}
{ }^P G_m^\phi\left(X_{S i}\right)={ }^P A^\phi ~X_{S i}^2+{ }^P B^\phi ~X_{S i}+{ }^P C^\phi
\label{parabola}
\end{aligned}
\end{equation}
where $^P A^\phi$, $^P B^\phi$ and $^P C^\phi$ are the parabola coefficients. Gibbs free energy curves for $\alpha$-Al and d-Si phases as well as their parabola approximations, for the numerical implementation, can be found in the supplementary material.
\subsection{Inter-diffusivity} 
As presented in \cite{Fujikawa1978}, the inter-diffusivity in Al-Si alloys is independent of the Si concentration. As a result, its value reduces to the impurity diffusion coefficient of Si in Al:
\begin{equation}
\begin{aligned}
{D{_{Si}^{\ast~\alpha}}^{}}= {D{_{Si~0}^{\ast~\alpha}}^{}}_{}  ~~exp \left (-\dfrac{^{Q_{D_{Si}^{\ast~\alpha}}}  }{RT} \right)
\label{e07a_inted_arrh}
\end{aligned}
\end{equation}
with the pre-exponential factor ${D{_{Si0}^{\ast~\alpha}}^{}}_{}=3.66 \times 10^{-6}~m^{2}s^{-1}$ and the activation energy ${Q_{D_{Si}^{\ast~\alpha}}}$=111 kJ.mol$^{-1}$ taken from  \cite{MANTINA20094102}. This inter-diffusivity ${D{_{Si}^{\ast~\alpha}}^{}}$ is indeed considered independent of Si concentration and only depends on temperature.
\subsection{Molar volume and eigen strain} \label{par_eigen_strain}
The molar volume for non magnetic materials can be expressed as-follows:
\begin{equation}
\begin{aligned}
V_{}^{\phi}(T)
=&
V_{0}^{\phi}~
exp
\left (
\int_{T_{0}}^{T}
3~\alpha_1^{\phi}~ dT
\right )
\label{e09_molar_volum_pur}
\end{aligned}
\end{equation}
where $V_{0}^{\phi}$  is the molar volume at the reference temperature $T_{0}$ (298 K) of the phase $\phi$ and $\alpha_1^{\phi}$ is its coefficient of linear thermal expansion. The composition dependence of the molar volume is expressed as Redlich-Kister formalism (\cite{HUANG2020101693}) where the data from \cite{HALLSTEDT2007292} are used. In the present case of coherent precipitate, the eigen strain $\varepsilon^{0}$ in the material comes from the lattice parameter misfit between the $\alpha$-Al and d-Si phases: 
\begin{equation}
\begin{aligned}
\varepsilon_{ij}^{0}
=&
\dfrac{a^{d}-a^{\alpha}}{a^{\alpha}}h(\eta)~\delta_{ij}
\label{e15_eigen_strain}
\end{aligned}
\end{equation}
where $\delta_{ij}$ is the Kronecker symbol, $a^{\alpha}$ and $a^{d}$ and are the lattice parameters.
\newline The model takes into account elastic energy. But in reality, mechanical dissipation takes place due to relaxation of the stress field around the precipitates as reported in \cite{FISCHER2015164}. To account for this effect, a correction factor ${f_{\varepsilon}}_{0}$ is applied to the eigen strain and is tuned as a fitting parameter in the model, so that ${\varepsilon_{0}}_{num}=f_{{\varepsilon}_{0}}~{\varepsilon_{0}}_{}$, where ${\varepsilon_{0}}_{num}$ is the eigen strain tensor put in the numerical model.
\subsection{Stiffness tensor} 
For the Al-Si system, to the best of our knowledge, there is no composition dependent CALPHAD-type evaluation of elastic constants available in the literature. In the present model, the elastic constants of pure Al are considered for the whole system. Su et al. \cite{Su201507} expressed the temperature dependent Young $\left(E^{\alpha}\right)$ and bulk $\left(B^{\alpha}\right)$ modulus of $\alpha$-Al single crystal reported in Table \ref{tab:calphad}. The stiffness tensor $C_{ijkl}$ in (Eq. \ref{eq:eq9}) is assembled from its three classical components calculated, based on $E^{\alpha}$ and $B^{\alpha}$, as follows.
\begin{equation}   
\begin{aligned}
C_{11}^{\alpha}=B^{\alpha}-\frac{4 E^{\alpha} B^{\alpha}}{E^{\alpha}-9 B^{\alpha}} ; ~~ C_{12}^{\alpha}=B^{\alpha}+\frac{2 E^{\alpha} B^{\alpha}}{E^{\alpha}-9 B^{\alpha}} ; \\ C_{44}^{\alpha}=-\frac{3 E^{\alpha} B^{\alpha}}{E^{\alpha}-9 B^{\alpha}}
\label{stiffness_tensor}
\end{aligned}
\end{equation}
Here, the elastic stiffness components of the $\alpha$ phase ($C_{11}^{\alpha}$, $C_{12}^{\alpha}$ and $C_{44}^{\alpha}$) are written in the shorthand Voigt's notation.
\subsection{Interfacial energy} 
The interfacial energy $\gamma$ plays an important role in the nucleation and coarsening processes. Its value is related to the coherency of the interface. Coherent interfaces have an interfacial energy range of $\left[ 0 - 0.2 \mathrm{~J}. \mathrm{m}^{-2} \right] $, semi-coherent of $\left[ 0.2 - 0.5 \mathrm{~J} . \mathrm{m}^{-2} \right] $ and incoherent of $\left[ 0.5-1 \mathrm{~J}. \mathrm{m}^{-2} \right] $ (\cite{PE92}). Several factors contribute to the interfacial energy, among which the strain energy. A high interfacial energy is expected for a high elastic strain energy and so corresponds to an incoherent interface. After some trials, and based on the assumption of an incoherent precipitation in AlSi10Mg alloy, a value of 1 $\mathrm{~J} . \mathrm{m}^{-2}$ was considered to conduct the present work. It should be noted that there are several ways to calculate the interfacial energy. Another approach for the determination of this parameter is given in the supplementary material.
\subsection{Enhanced diffusion by quenched-in excess vacancies} \label{enhanced_diffusion_sec}
This is the dominant mechanism for the diffusion of solvent matrix atoms and substitutional solute ones in metals (\cite{Meh07}). In this mechanism, the solute or tracer atom jumps into a neighboring vacancy. The expression of the impurity diffusion coefficient ${D}_{Si}^{*~\alpha}$ in Eq. \ref{e07a_inted_arrh} can be written in terms of vacancy equilibrium site fraction $X_{Va}^{e}$ for a cubic Bravais lattice matrix as introduced in \cite{Meh07}:
\begin{equation}
\begin{aligned}
{D}_{Si}^{*~\alpha} = 
f_{2}~{(a^{\alpha})}^{2}~\omega_{2}~X_{Va}^{e}
\left (
\dfrac{G_{m}^{b}}{RT}
\right)
\label{e01_diffu_coeff}
\end{aligned}
\end{equation}
where $f_{2}$ is the impurity correlation factor which accounts for the deviation of the random walk behavior, $\omega_{2}$ is the vacancy impurity exchange rate, and $G_{m}^{b}$ is the molar Gibbs energy of vacancy-solute binding. The equilibrium vacancy site fraction depends on the temperature:
\begin{equation}
\begin{aligned}
X_{Va}^{e} = 
exp
\left (
\dfrac{-{G_{m}^{f}}_{Va}}{RT}
\right)
\label{e02_vacan_conce}
\end{aligned}
\end{equation}
where $G_{m}^{f_{Va}}$ is the molar Gibbs energy of vacancy formation.
However, there is no clear value of the impurity correlation factor ($f_{2}$) for the diffusion of Si in $\alpha-Al$ within the literature. 
 An alternative is to consider that vacancies are created or annihilated by sources or sinks which are dislocations, grain boundaries and incoherent precipitate interfaces. If the material undergoes a rapid cooling, the vacancies do not have enough time to reach the sink and are entrapped. In our work, those quenched-in excess vacancies enhance the solute diffusion and Eq. \ref{e07a_inted_arrh} is accordingly modified:
\begin{equation}
\begin{aligned}
{D}_{Si}^{*~\alpha} = 
\left (\dfrac{X_{Va}}{X_{Va}^{e}} \right)
{D}_{Si_{~0}}^{*~\alpha}~
exp
\left (
-\dfrac{Q_{{D}_{Si}^{*~\alpha}}}{RT}
\right)
\label{e03_diffu_effec}
\end{aligned}
\end{equation}
where $X_{Va}$ is the fraction of excess vacancies.
\newline The evolution of the excess vacancies can be represented by an exponential law from \cite{Falahati2010}: 
\begin{equation}
\begin{aligned}
X_{Va}(t+dt) =  X_{Va}(t)+
\left [1-
exp
\left (-
\dfrac{D_{Va}}{{L_{Va}}^{2}}dt
 \right)
 \right ] [X_{Va}^{e}(T(t))-X_{Va}(t)]
\label{e04_vacan_gover}
\end{aligned}
\end{equation}
where $L_{Va}$ is the mean diffusion distance of the vacancies to the sources and sinks. The vacancy diffusion coefficient $D_{Va}$ is expressed as an effective coefficient between the impurity diffusion of Si in Al ${D}_{Si}^{*~\alpha}$ and the self-diffusion of Al ${D}_{Al}^{*~\alpha}$:
\begin{equation}
\begin{aligned}
D_{Va} = 
\dfrac{X_{Va}}{X_{Va}^{e}}~
(
X_{Si}~
{D}_{Si}^{*~\alpha}
+(1-X_{Si})~
{D}_{Al}^{*~\alpha}
)
\label{e05_vacan_diffu}
\end{aligned}
\end{equation}
The effect of cooling rate on the vacancy site fraction is detailed in the supplementary material.
\subsection{CALPHAD data} 
The CALPHAD parameters used to compute quantities described above are summed up in Table \ref{tab:calphad}. 
\begin{table*}[htbp]
\centering
\caption{Parameters of the binary system Al-Si, each parameter is defined by a function: a + $b T+c T^{2}+d T^{-2}+e T \ln T+f T^{3}+g T^{-1}+h T^{-9}$}
\label{tab:calphad}
\begin{tabular}{llllll}
\toprule				
\hline							
Parameter	&	a	&	b	&	c	&	d	&	Ref	\\
\hline			
\midrule											
$\!^{o}G_{Al}^{\alpha}$$\!^{(1)}$ 	& $-7976.15$	&	$137.093038$	&	$-1.884662\times 10^{-3}$	&		&	\cite{ADR98}	\\
$\!^{o}G_{Al}^{\alpha}$$\!^{(2)}$ 	&	$-11276.24$ 	&	$233.04844$	&	$18.531982\times 10^{-3}$	&		&		\\
$\!^{o}G_{Al}^{\alpha}$$\!^{(3)}$	&	$-11278.378$	&	$188.6841536$	&		&		&		\\
											
$\!^{o}G_{Si}^{\alpha}$ - $\!^{o}G_{Al}^{\alpha}$	&	$51000.00$	&	$-21.8$	&		&		&		\\
$\!^{o}\Omega_{AlSi} ^{\alpha}$	&	$-3143.78$ 	&	$0.39297$ 	&		&		&		\\
$\!^{o}G_{Si}^{d}$$\!^{(4)}$		&$-8162.609$	&	$137.236859$	&	$-1.912904\times 10^{-3}$	&		&		\\
$\!^{o}G_{Si}^{d}$$\!^{(5)}$	& 	$-9457.642$	&	$167.281367$	&	$18.531982\times 10^{-3}$	&		&		\\
$\!^{o}G_{Al}^{d}$   - $\!^{o}G_{Al}^{\alpha}$	&		&	$30$	&		&		&		\\
$\!^{o} \Omega_{AlSi}^{d}$\vspace{5pt}	&	$113246.16$	&	$-97.34$	&		&		&		\\

$V_{0 \ Al}^{\alpha}$	&	$9.77430 \times 10^{-6}$	&		&		&		&	\cite{HALLSTEDT2007292}	\\
$V_{0 \ Si}^{\alpha}$	&	$9.2 \times 10^{-6}$	&		&		&		&		\\
$V_{0 \ Si}^{d}$	&	$12.0588 \times 10^{-6}$	&		&		&		&		\\
$3\alpha_{Si}^{\alpha}$ 	&	$1.12466 \times 10^{-5}$	&	$8.92959 \times 10^{-9}$	&	$-3.09709 \times 10^{-12}$	&	$-0.251151$	&		\\
$3\alpha_{Al}^{\alpha}$	&	$6.91213 \times 10^{-5}$	&		&	$4.86802 \times 10^{-11}$	&	$-0.413484$	&		\\
$3\alpha_{Si}^{d}$\vspace{5pt}	&	$8.56511 \times 10^{-6}$	&	$6.85005 \times 10^{-9}$	&	$-2.35401 \times 10^{-12}$	&	$-0.191371$	&		\\
											
$E^{\alpha}$	&	$8.030 \times 10^{9}$	&	$-1.319 \times 10^{7}$	&	$-6.815 \times 10^{4}$	&		&	\cite{Su201507}	\\
$B^{\alpha}$ \vspace{5pt}	&	$7.028 \times 10^{9}$	&	$2.114 \times 10^{6}$	&	$-4.903 \times 10^{3}$	&		&		\\
\midrule			
\hline											
Parameter	&	e	&	f	&	g	&	h	&	Ref	\\
\hline			
\midrule											
$\!^{o}G_{Al}^{\alpha}$$\!^{(1)}$ 	&	$24.3671976$	&	$-0.877664\times 10^{-6}$	&	$74092$	&		&	\cite{ADR98}	\\
$\!^{o}G_{Al}^{\alpha}$$\!^{(2)}$	&	$-38.5844296$	&	$-5.764227\times 10^{-6}$	&	$74092$	&		&		\\
$\!^{o}G_{Al}^{\alpha}$$\!^{(3)}$	&	$-31.748192$	&		&		&	$-1230.524\times 10^{25}$	&		\\

$\!^{o}G_{Si}^{d}$$\!^{(4)}$	&	$-22.8317533$	&	$-0.003552\times 10^{-6}$	&	$176667$			&		\\
$\!^{o}G_{Si}^{d}$$\!^{(5)}$	&	$-27.196$	&		&			$-420.369\times 10^{28}$	&		 \\
\bottomrule
\multicolumn{6}{@{}l}{\vspace{-7pt}} \\
\hline	
\multicolumn{6}{@{}l}{ \footnotesize  
$^{(1)}$  for 298.15$<$T$\prec$700 K ; $^{(2)}$ for 700$<$T$\prec$933.47 K ; $^{(3)}$ for 933.47$<$T$\prec$2900 K}  \\
	
\multicolumn{6}{@{}l}{ \footnotesize  
$^{(4)}$  for 298.15$<$T$\prec$1687 K ; $^{(5)}$ for 1687$<$T$\prec$3600 K }

\end{tabular}
\end{table*}  
\section{Implementation of the KKS model}  \label{gov_eq}
\subsection{Dimensionless kinetic equations }
For numerical purposes, dimensionless variables are used so that Eqs. \ref{e01_cahn_hilli} and \ref{e02_allen_cahn} are transformed into dimensionless forms. This transformation is mandatory to guarantee efficient computing as justified by Hu et al. \cite{HU2007303}. The equation then read:
\begin{equation}
\begin{aligned}
\dfrac{\partial X_{Si}}{\partial t^{\star}}  =
\tilde{D}^{\star}
\nabla^{\star} . \left [
  \dfrac{{f_{0}}_{X_{Si}\eta}}{{f_{0}}_{X_{Si}X_{Si}}} \nabla^{\star} \eta +\nabla^{\star}  X_{Si}
 \right ]
\label{e01_cahn_norma}
\end{aligned}
\end{equation}
\begin{equation}
\begin{aligned}
\dfrac{\partial \eta}{\partial t^{\star}}
=& -L^{\star}
\left [
f_{0 \ \eta}^{\star}
-{\kappa^{\star}}^{2}{\nabla^{\star}}^{2}\eta
+{e_{el}^{\star}}_{\eta}
\right ]
\label{e02_allen_norma} 
\end{aligned}
\end{equation}
It should be noted that in order to increase the concision and clarity of the mathematical formulations, the notation  $\tilde{D}_{}$ is hereafter used instead of ${D{_{Si}^{\ast~\alpha}}^{}}$.
\\ The derivatives read:
\begin{equation}
\begin{aligned} 
\frac{f_{0 \ X_{Si}\eta}}{f_{0\ {X_{Si}} X_{S i}}}=\frac{\partial^2 f_0 / \partial X_{Si} \partial \eta}{\partial^2 f_0 / \partial X_{S i}{ }^2}=\frac{d h}{d \eta}\left(X_{S i}^\alpha-X_{S i}^d\right)
\label{e01_cahn_norma_deriv}
\end{aligned}
\end{equation}
and:
\begin{equation}
\begin{aligned}
f_{0 \ \eta}^{\star}=\frac{\partial f_0^{\star}}{\partial\eta}=-\frac{d h}{d \eta}\left({f^\alpha}^{\star}-{f^d}^{\star}-\left(X_{S i}^\alpha-X_{S i}^d\right) \frac{\partial {f^\alpha}^{\star}}{\partial X_{S i}^\alpha}\right)-w^{\star} \frac{d g}{d \eta}
\end{aligned}
\end{equation}
with the adimensional parameters : $\tilde{D}^{\star}= \frac{\tilde{D}}{\tilde{D}(t=0)_{}},~ t^{\star}=\frac{\tilde{D} }{\Delta x}t,~ \nabla^{\star}=\frac{\nabla}{\Delta x},~ L^{\star}=\frac{L C_{44} \Delta x^{2}}{\tilde{D}},~ \kappa^{\star 2}=\frac{\kappa^{2}}{C_{44} \Delta x^{2}}, ~w^{\star}=\frac{w}{C_{44}},~ f^{\phi^{\star}}=\frac{f^{\phi}}{C_{44} V_{m}},~ e_{el_ \eta}^{\star}=\frac{e_{e l_ \eta}}{C_{44}}~where~ e_{e l_ \eta}=\frac{d e_{e l}}{d\eta}
$
\newline \newline As justified in \cite{Fetni2021_COMPLAS}, the first order semi-implicit Fourier spectral scheme is used to discretize the normalized CH and AC equations (Eqs.  \ref{e01_cahn_norma} and \ref{e02_allen_norma}) as follows:
\begin{equation}
\begin{aligned}
\dfrac{\{ X_{Si} \}_{k}^{n+1} -\{ X_{Si} \}_{k}^{n}}{\Delta t ^{\star} }
=&
\tilde{D}^{\star}i{k}.
\left \{
\dfrac{{f_{0}}_{X_{Si}\eta}^{}}{{f_{0}}_{X_{Si}X_{Si}}^{}}
[i{k'} \{\eta \}_{k'}^{n}]_{r}
\right \}_{k}^{n}
-\tilde{D}^{\star}k^{2}\{ X_{Si} \}_{k}^{n+1}
\label{e03_cahn_norma_discr} 
\end{aligned}
\end{equation}
\begin{equation}
\begin{aligned}
\dfrac{\{ \eta \}_{k}^{n+1} -\{ \eta \}_{k}^{n}}{\Delta t ^{\star}}
=& 
-L^{\star}
\left [
\{ {f_{0}^{\star}}_{\eta} \}_{k}^{n}
+
{\kappa^{\star}}^{2}k^{2}
\{ \eta \}_{k}^{n+1}
+\{{ e_{el}}_{\eta}^{\star}\}^{n}_{k}
\right ]
\label{e04_allen_norma_discr} 
\end{aligned}
\end{equation}
with $\left\{~.~ \right\}_{k}$ the Fourier transform and $ \left[~.~\right]_{r}$ the inverse Fourier transform. \textit{k'} is used to distinguish different Fourier transformations.
\newline The stress equilibrium in Eq. \ref{eq:eq10} is solved at each time step by the Fourier spectral method as well following the developments in \cite{MOULINEC199869} and \cite{Bin17}.
\begin{equation}
\begin{aligned}
\{ \varepsilon_{kl} \}_{k_f}
=& 
\{ \varepsilon_{kl}^{0} \}_{k_f}
-\Gamma_{khij} \{ \sigma_{ij}^{0} \}_{k_f}
\label{e05_elast_defor}
\end{aligned}
\end{equation}
where $k_f$ indicates the Fourier transform and  $\Gamma_{khij}$ is the Green operator in the Fourier space:
\begin{equation}
\begin{aligned}
\Gamma_{khij}
=& 
\dfrac{1}{4}
[
N_{hi}k_{j}k_{k}+N_{ki}k_{j}k_{h}+N_{hj}k_{i}k_{k}+N_{kj}k_{i}k_{h}
]
\label{e06_green_tenso}
\end{aligned}
\end{equation}
with $N_{ki} = N_{ik}^{-1}$ and $N_{ik}=C_{ijkl}k_{j}k_{l}$
\subsection{Adaptive time stepping scheme}
The implementation of an adaptive time stepping strategy is mandatory for phase-field simulations due to the big difference between time scales needed to accurately model the complex physical phenomena. Indeed, large steps could be allowed in the time intervals where there is no noticeable variation of the computed solution, in order to speed up the simulation. However, the solution becomes sensitive to the time step when physical changes occur. 
The adaptive scheme applied in this work is derived from those proposed in \cite{GUILLENGONZALEZ2014821}. The time step is indeed dynamically updated in function of the evolution of the residual of the discrete energy law. We first tested this strategy introduced in \cite{Fetni2021_COMPLAS} to study the spinodal decompositions of binary alloys, governed by CH equation. 
\newline The residual of the KKS formulation is obtained by deriving Eq. \ref{eq:eq1} with respect to time:
\begin{equation}
\begin{aligned}
RES_{KKS}=\frac{d \mathcal{F}_{KKS}}{d t} &-\int_V\left[- M\left|\nabla \frac{\delta \mathcal{F}_{KKS}}{\delta X_{Si}}\right|^2-\frac{1}{L} \eta_t^2+f_{0 T} T_t \right] d V
\label{eq:res_kks}
\end{aligned}
\end{equation}
where $\eta_t$ and $T_t$ are the shorthand notations of the derivatives of $\eta$ and \textit{T} with respect to time, while $f_{0 T}$ is the derivative of $f_{0}$ with respect to T. Note that we consider the dependance of the total Helmholtz free energy $\mathcal{F}_{KKS}$ and the homogeneous free energy density $f_{0}$ on $X_{Si}$, $\eta$ and \textit{T}.
The residual $RES_{KKS}$ is computed for the (n+1) time-step at $t = t^{n+1/2}$ (shorthand notation of  $\frac{ t^{n}+t^{n+1}}{2}$).
The implementation of the adaptive time stepping scheme is illustrated in Algorithm \ref{algorithm_1}.
Here \textit{resmin}, \textit{resmax }, $\textit{iter}_{\textit{max}}$ and $\theta$ are the parameters of the adaptive scheme. It is based on an iterative loop and the energy residual is controlled at each time step to ensure that it remains bounded between the two predefined limits \textit{resmin} and \textit{resmax}. By this way, the accuracy of the scheme is guaranteed. The time step is progressively increased when the residual becomes below \textit{resmin}, while the scheme requires a number of iterations (up to $\textit{iter}_{\textit{max}}$) to identify the accurate time step and recomputes the solution when \textit{resmax} is exceeded. If the convergence is not reached, then the simulation is stopped and further action should be taken before restarting the simulation (fine tuning the model parameters including those of the time stepping scheme).
\begin{scriptsize}
\begin{algorithm}[htbp]
    \caption{Adaptive time stepping scheme}\label{algorithm_1}
    \begin{algorithmic}
        \State  $\mathrm{initialization}~~ \{\eta,X_{Si}\}$
        \For{steps = $1$ to $N$ (Total number of time steps)}
					\State iter=1
									\While{$(iter \leq iter_{max})$}
											\State compute $\eta^{n+1}$ from  $\eta^{n}~$ $\leftarrow$ Eq. \ref{e04_allen_norma_discr}
											\State compute $X_{Si}^{n+1}$ from  $X_{Si}^{n}~$  and $\eta^{n}$ $\leftarrow$ Eq. \ref{e03_cahn_norma_discr} 
											\State compute $RES_{KKS}$ at $t = t^{n+1/2}$ ~$\leftarrow$ Eq. \ref{eq:res_kks}
											\If{$\left|RES_{KKS}\right| > \textit{resmax}$} 
													\State take $\Delta t^{n}= \frac{\Delta t^{n}}{\theta} $
													\State $iter+=1$
													\State \textbf{repeat} the actual step 
											\Else
													\If{$\left|RES_{KKS}\right| < \textit{resmin}$}
															\State take $\Delta t^{n}= {\theta}.{\Delta t^{n}}$
															\State \textbf{\textit{break}} the while loop and \textbf{go to} next step
													\Else 
														 \State nothing to do 
                                    \State{$\textit{resmin} < \left| RES_{KKS} \right| < \textit{resmax}$}

														 \State \textbf{\textit{break}} the while loop and \textbf{go to} next step
													\EndIf
											\EndIf 
											\If{$iter > iter_{max}$}
											\State \textbf{\textit{Exit}} 
											\EndIf
									\EndWhile
							\EndFor
	\end{algorithmic}
\end{algorithm}
\end{scriptsize}
\section{Application of the implemented KKS model} \label{results}
To demonstrate the application of the proposed KKS phase-field model, we propose hereafter to simulate a DSC anisothermal heat load, to explain the observed peaks and to predict microstructural evolution and thermo-physical properties.
\subsection{Microstructure initialization} \label{micro_initialization}
\begin{figure*}[htbp]
\centering
{\includegraphics[scale=0.4]{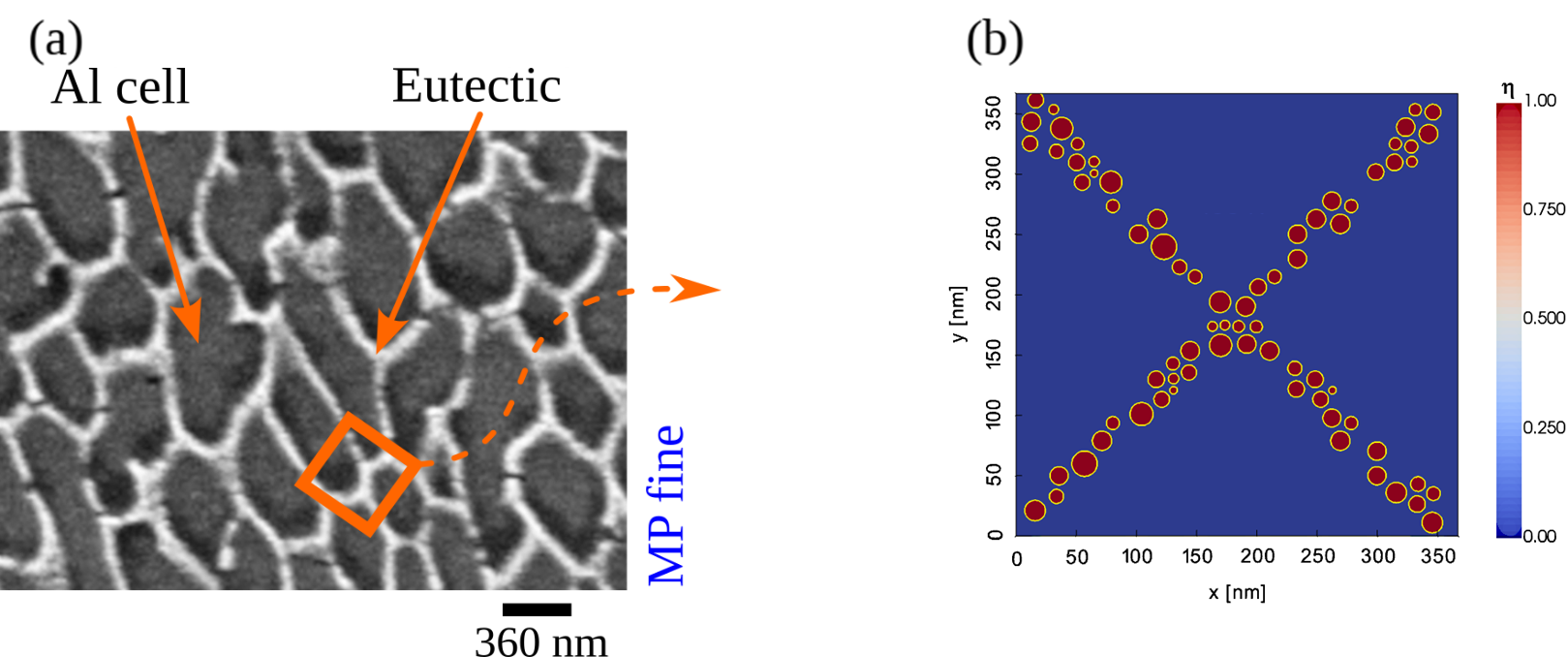}}
\caption{(a) Experimental micrograph of the MP fine (b) Corresponding phase-field 2D domain.}     
\label{fig:vacancy}
\end{figure*}

It is reminded that the microstructure of AlSi10Mg LPBF is mainly composed of MP fine. Thus, this zone is only taken into account in the following phase-field simulations. The case study concerns a sample manufactured with a $35^{\circ} \mathrm{C}$ preheated platform and exhibits the following features: a 360 $\times$ 360 nm cell size, a molar volume fraction and equivalent radius of d-Si precipitates of $V^{d_0}=0.083$ and $r^d=7 \mathrm{~nm}$ respectively and a Si solute level in the $\alpha$-Al matrix of $X_{S i}^{\alpha_0}=0.025$. For the simulation, a 2D domain of $710 \times 710 \mathrm{~nm}$ is used to represent four MP fine cells (periodic microstructure) with a quadruple point within the eutectic mixture as shown in Fig. \ref{fig:vacancy} (a) and (b). 
 Such a simulation domain allows obtaining significant averages of some key quantities (in particular for Si molar fraction in the $\alpha$-Al matrix, phase fractions of d-Si precipitates and their specific surfaces). The d-Si precipitates in the eutectic mixture were created from a normal distributed packing of discs of mean radius $r^{d}$ and standard deviation $0.1 r^{d}$. For that purpose, the disc packing was created using the solid dynamics physic module of Blender software (\cite{Hes10}).
\subsection{Post-processing of phase-field results to obtain DSC curves}
Direct comparison between phase-field modeling and DSC experiment can be made by computing a DSC curve. According to \cite{BOTTGER20096784}, the integral heat flux per unit of sample mass $\dot{q}_W$ reads:
\begin{equation}
\begin{aligned}
\dot{q}_W=\frac{1}{V d t} \int_V \sum_\phi \frac{1}{u_m^\phi} d\left(H_m^\phi \eta^\phi\right) d V
\label{eq:heat_flux}
\end{aligned}  
\end{equation}
where $V$ is the volume of the simulation domain, $H_m^\phi$ is the molar enthalpy of the phase $\phi, \eta^\phi$ is the order parameter of the phase $\phi\left(\eta^d=\eta~ where~ \eta >0.5, \eta^\alpha=\eta~ where~ \eta<0.5\right)$ and $d\left(H_m^\phi \eta_\phi\right)$ can be expressed as follows:
\begin{equation}
\begin{aligned}
d\left(H_m^\phi \eta^\phi\right)=\eta^\phi\left(C_p^\phi d T+\sum_i \frac{\partial H_m^\phi}{\partial X_i} d X_i\right)+H_m^\phi d \eta^\phi+\gamma_H~ d A^d
\label{eq:d_H_m}
\end{aligned}  
\end{equation}
where $C_p^\phi$ is the heat capacity of the phase $\phi=\left\{d-Si,\alpha \right\}$ and $\gamma_H$ the interfacial enthalpy. $C_p^\phi$ is responsible for the DSC baseline, the second term corresponds to the heat of diffusion and the third term $H_m^\phi d \eta^\phi$ is the heat of precipitation or dissolution during the phase transformation. The last term $\gamma_H ~dA^d$ is added to the original expression to take into account the heat released during the coarsening due to the reduction of the precipitate area $d A^d$. The heat capacity of phase $\phi, C_p^\phi$, is computed from CALPHAD data:
\begin{equation}
\begin{aligned}
C_p^\phi=-T\left(\frac{\partial^2 G_m^\phi}{\partial T^2}\right)_{P, X_i}
\label{eq:heat_capacity_phi}
\end{aligned}  
\end{equation}
Eq. \ref{eq:heat_flux} can be approximated by taking average values over the whole domain, noted $\left\langle~ \right\rangle$:
\begin{equation}
\begin{aligned}
\dot{q}_W=\frac{1}{V d t} \int_V \sum_\phi \frac{1}{u_m^\phi} d\left(\left\langle H_m^\phi\right\rangle\left\langle\eta^\phi\right\rangle\right) d V
\label{eq:heat_capacity_phi_}
\end{aligned}  
\end{equation}
The apparent heat capacity of the alloy is computed by:
\begin{equation}
\begin{aligned}
C_p^*=\frac{\dot{q}_W}{\dot{T}}
\label{eq:apparent_heat_capacity_phi}  
\end{aligned}  
\end{equation}
The interfacial enthalpy $\gamma_H$ is estimated from the interfacial energy $\gamma$ and the interfacial entropy $\gamma_S$ from $\gamma=\gamma_H-T ~\gamma_S$. According to \cite{BOYD19711101}, the interfacial entropy is approximately $0.5-1 \times 10^{-3} \mathrm{~J}.\mathrm{m}^{-2}.\mathrm{K}^{-1}$, which gives for the temperature of interest $\left[400-800 \mathrm{~K}\right]$ an average value $\gamma_H=1.5~ \mathrm{J.m}{ }^{-2}$.
\subsection{Microstructure evolution of a representative AlSi10Mg LPBF Cell under anisothermal ageing (to simulate a performed DSC test)} \label{case_study}
\begin{figure*}[htbp]
\includegraphics{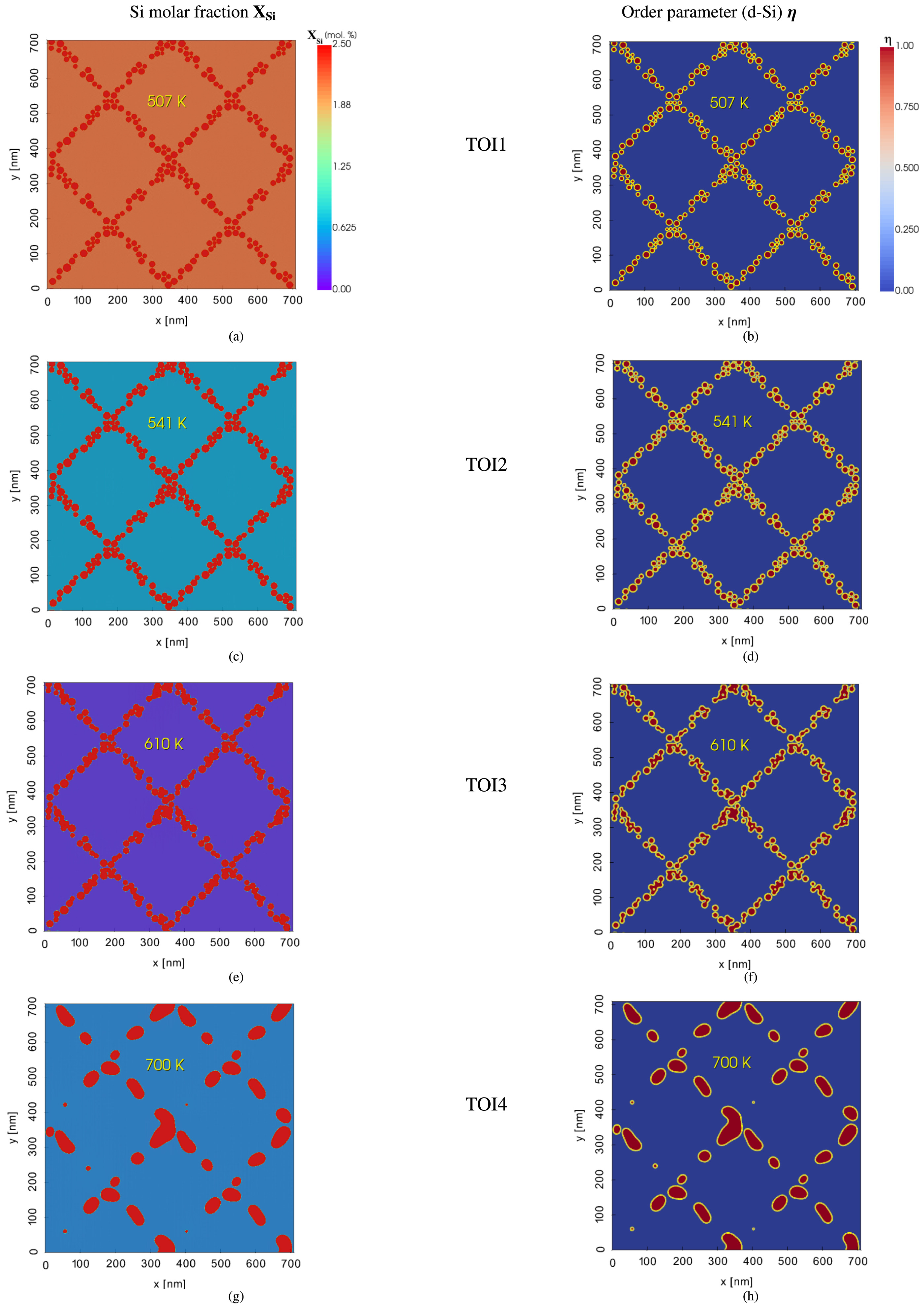}
\caption{2D maps of the Si molar fraction ($X_{Si}$, left) and the d-Si phase fraction/order parameter ($\eta$, right) for different temperatures: \textit{T}=507 K [(a)-(b)], \textit{T}=541 K [(c)-(d)], \textit{T}=610 K [(e)-(f)] and \textit{T} = 700 K [(g)-(h)]. The yellow contour lines represent the iso-values for $\eta$=0.1 and $\eta$=0.9 showing the interface thickness between the d-Si precipitates and the $\alpha$-Al matrix. A movie showing $X_{Si}$ evolution is provided in the supplementary material ; the Readers are invited to consult the electronic version of this paper.}
\label{fig:2D_maps}

\end{figure*}

\begin{figure*}[htbp]
    \centering
    \begin{tabularx}{\linewidth}{@{} X X @{}}
        \begin{subfigure}{0.45\textwidth}
            \centering
            \includegraphics[width=\linewidth]{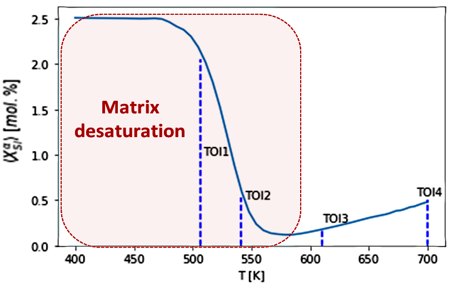}
            \caption{Evolution of $\langle X_{Si}^{\alpha} \rangle$}
            \label{fig:sub1}
        \end{subfigure}%
        &
        \begin{subfigure}{0.45\textwidth}
            \centering
            \includegraphics[width=\linewidth]{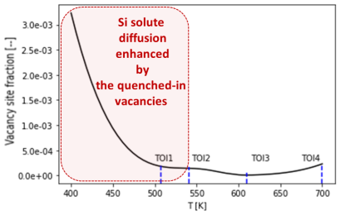}
            \caption{Evolution of vacancy site fraction}
            \label{fig:sub2}
        \end{subfigure}%
        \\
        \begin{subfigure}{0.45\textwidth}
            \centering
            \includegraphics[width=\linewidth]{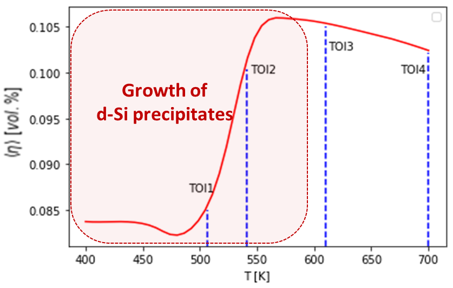}
            \caption{Evolution of $\langle \eta \rangle$}
            \label{fig:sub3}
        \end{subfigure}%
        &
        \begin{subfigure}{0.45\textwidth}
            \centering
            \includegraphics[width=\linewidth]{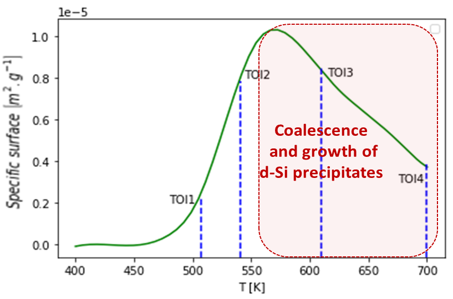}
            \caption{Evolution of d-Si specific surface}
            \label{fig:sub4}
        \end{subfigure}%
    \end{tabularx}
    \caption{Evolution of: (a) the average Si molar fraction in the $\alpha$-Al matrix $\langle X_{Si}^{\alpha} \rangle$, (b) vacancy site fraction, (c) d-Si phase fraction $\langle \eta \rangle$, and (d) d-Si specific surface with respect to the temperature for a heating rate of 20 K.min$^{-1}$.}
    \label{fig:quantity_evolution}
\end{figure*}

A heating rate of 20 K/min was applied to the 2D domain of 710 $\times$ 710 ${nm}^{}$ starting from $\mathrm{\texttt{T}}=400 \mathrm{~K}$. The interfacial energy $\gamma$ is set to $1 \mathrm{~J} . \mathrm{m}^{-2}$. The interface thickness is linked to the grid size $\Delta x=\Delta y=0.5 \mathrm{~nm}$ ; $2 \lambda=5 \Delta x$. Indeed, after some trials, and based on \cite{HU2007303}, it was found that the range of $\Delta x$ between 0.33 and 0.5 allows stable simulations.
\newline Accurate model parameter identification was mandatory to represent the correct transformation kinetics with the model. For that, in a similar way to the introduction of ${f_{\varepsilon}}_0$ (c.f. paragraph \ref{par_eigen_strain}), correction coefficients for the interface mobility, diffusivity, interfacial energy and kinetic coefficient of the interface were introduced : ${f_{M_{\eta}}}$, $f_{\widetilde{D}}$, ${f_{\gamma_{}}}$ and $f_L$ respectively. A sensitivity analysis was conducted to identify the optimum set of parameters, to ensure that the computed DSC result curve best matches the experimental one.
It is worth noting that these parameters, once identified, are fixed for the whole numerical experiment and used to correct the related physical quantities (interface mobility $M\eta$, diffusivity $\widetilde{D}$, interfacial energy $\gamma$ and interface kinetic coefficient $L$, respectively) by the same factor during the entire simulation.
The set of parameters includes three subsets: grid spacing, correction coefficients for material parameters and adaptive time stepping scheme.
They are gathered in Table \ref{tab:ident_param}. Details about the conducted sensitivity analysis to identify the material and numerical parameters and explain the sensitivity of the solution are presented in the supplementary material. 
Although the correction coefficient ${f_{\varepsilon}}_{0}$
 for the elastic energy was set to a small value of 0.01 in Table \ref{tab:ident_param}, this does not imply that the elastic energy was neglected in the calculations. Neglecting the elastic energy, especially for LPBF-processed materials with significant lattice mismatches, could lead to inaccurate predictions of the microstructure evolution. Our numerical experiments and sensitivity analysis identified a stable range for ${f_{\varepsilon}}_{0}$ between 0.001 and 0.1. Refinement targeted the identification of the experimental DSC peaks, with up to 10\% of the elastic energy ($e_{el}$) allowing for stable computations. Higher values led to numerical artifacts, suggesting an excessive consideration of the elastic energy. Quantifying the elastic energy contribution is complex, involving phenomena such as creep, plastic deformation, incoherent interfaces...\cite{AAGESEN201710,Hu2004} and modeling aspects like the absence of the third dimension \cite{FETNI2021109661}. Thus, ${f_{\varepsilon}}_{0}$ was included as a fitting parameter to account for these factors in a simplified manner.
\begin{table*}[htbp]
\centering
\caption{Identified spatio-temporal and material parameters to model microstructure evolution of the representative AlSi10Mg LPBF cell during anisothermal ageing (heating rate of 20 K/min).}
\label{tab:ident_param} 
\begin{tabular}{|c|c|c|c|c|c|c|c|c|c|c|}
\hline
\textbf{Grid spacing} & \multicolumn{5}{c|}{\textbf{Time stepping scheme parameters}} & \multicolumn{5}{c|}{\textbf{Identified coefficients for material parameters}} \\
\hline
$\Delta x$, $\Delta y$ [nm] & $\Delta t^*$ & \textit{resmin} & \textit{resmax} & $\theta$ & $iter_{max}$ & $f_{M_\eta}$ & $f_{\widetilde{D}}$ & $f_\gamma$ & $f_{\varepsilon_0}$ & $f_L$ \\
\hline
0.5 & 250 & $10^{-4}$ & $10^{-1}$ & 1.25 & 15 & 1 & 1 & 2.15 & 0.01 & 0.01 \\
\hline
\end{tabular}
\end{table*}
 The evolution of the microstructure of AlSi10Mg LBPF under the simulated DSC test is shown in Fig. \ref{fig:2D_maps}. The color maps of the Si molar fraction and order parameter are shown for different heating times and maximum temperatures identified as temperatures of interest (TOIs): 507K, 541K, 610K and 700K. 
The corresponding mean value of Si molar fraction in $\alpha$ phase $\langle X_{Si}^{\alpha}\rangle$, mean volume fraction of d-Si precipitates $\langle \eta \rangle$ computed from the order parameter, specific surface of d-Si precipitate and vacancy site fraction are plotted in Figs. \ref{fig:quantity_evolution} (a), (b), (c), and (d) respectively. Here, three main stages can be distinguished.
The first stage is characterized by a high decrease of the Si molar fraction within the $\alpha$-Al matrix, associated with an increase of the d-Si phase fraction. This trend becomes clear at TOI1 and still steeps between TOI1 and TOI2. Those rapid variations are driven by the excess vacancies which enhance the diffusion of Si in this temperature range.
 By the end of this stage, the matrix is desaturated and the mechanism of growth of d-Si precipitates is triggered. TOI3 could be associated with a transient state between the precipitates growth phase and their coalescence. The third stage, linked to the coalescence mechanism, reduces d-Si precipitates specific surface (TOI4). The initial network is clearly segregated. Later, the d-Si precipitates dissolve due to the increase of the Si solubility as given by the Al-Si phase diagram (Fig. \ref{fig:flowchart}).

\par
\begin{figure}[htbp]
\centering
\subcaptionbox{ \label{}}
{\includegraphics[scale=0.65]{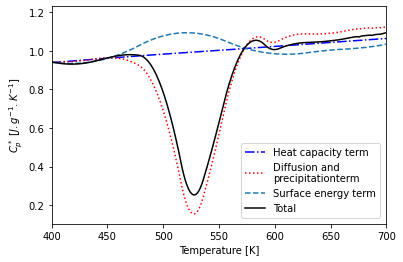}}

\subcaptionbox{ \label{}}
{\includegraphics[scale=0.65]{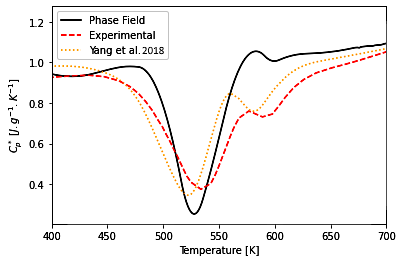}}
\caption{(a) Simulated apparent heat capacity curve. The signal is represented with its different terms. (b) Simulated heat capacity curve versus experiments (this work and those obtained from \cite{yang_deibler_bradley_stefan_carroll_2018}).}
\label{fig:C_p}
\end{figure}

 From the curves (Figs. \ref{fig:quantity_evolution} (a), (b), (c)) and Eqs. \ref{eq:heat_capacity_phi_} and \ref{eq:apparent_heat_capacity_phi}, the apparent heat capacity is plotted in Fig. \ref{fig:C_p} (a) with its different constitutive terms and is compared to DSC experiments in Fig. \ref{fig:C_p} (b).
The two peaks of the simulated curve are in agreement with the experimental ones in term of position and shape, but their amplitudes are a bit different especially for the second peak.
The first peak corresponds to the desaturation of $\alpha$-Al matrix from Si, while the second one is linked to the heat released during the coalescence of d-Si precipitates. This slight discrepancy can be ascribed to some approximations that have been introduced regarding both the experimental DSC measurements and the CALPHAD parameters. From an experimental viewpoint, the DSC device had been calibrated based on the latent heat of fusion of certified samples of pure substances, following standard practices (\cite{Höhne_Hemminger_Flammersheim_2003}). Calibration based directly on the specific heat of a reference sample as close as possible to AlSi10Mg LPBF might improve the agreement between experiments and prediction (\cite{BOYD19711101}). For the CALPHAD parameters, elastic constants of pure Al were considered due to the lack of data within literature, which could result in difference between simulated and measured heat capacity curves especially at high temperatures.
\par The simulated microstructure evolution is in accordance with the experimental analysis conducted in \cite{LI201574}. When studying the microstructure evolution of Al12Si
 (additively manufactured by LPBF) during heat treatment (at 500 $^\circ$C for up to 4 h), these authors reported that the mechanism of growth of Si particles, at longer times,
  is controlled by the combination of two phenomena : the Ostwald ripening on the one hand and the coalescence with the adjacent small precipitates on the other hand.
   Indeed, the noticeable growth of d-Si precipitates at temperatures above 500$^\circ$C is ensured by consuming small adjacent precipitates, as shown in Fig. \ref{fig:quantity_evolution}. The developed phase-field model can thus explain the governing mechanisms occurring during the solid transformation of the Al-Si system during heat treatment. The interface mobility $M_\eta$  plays a crucial role in determining the rate-controlling mechanism for the microstructural evolution process.
At high temperatures (end of the simulation), the interface mobility $M_\eta$ is very high, then the driving force at the interface $\Delta \mu_{S i}^i$ would go to 0 . In such case,  the Si molar fraction profile at the d-Si precipitate interface ($X_{S i}^{\alpha i}$) could be approximated: $X_{S i}^{\alpha i} \approx X_{S i}^{\alpha e}$. As a consequence, there is a local equilibrium at the interface and the interface velocity would be as fast as diffusion allows. The reaction is said diffusion controlled. Conversely, at lower temperatures (beginning of the simulation), $M_\eta$ is close to 0, then a maximal diffusion potential is required to drive the interface reaction, then $X_{S i}^{\alpha i} \approx X_{S i}^{\alpha 0}$. Thus, the reaction is said to be interface controlled. Between those 2 extreme cases, the reaction is mixed control. Furthermore, from the knowledge of the microstructure corresponding to specific LPBF and post-treatment process conditions, one can derive mechanical properties such as yield strength and uniform elongation, by using analytical formulas \cite{SANTOSMACIAS2020231}  \cite{CHEN20135877}. Such ability paves the way towards microstructure design to reach desired properties in LPBF Al alloys.
\section{Conclusions}
\par In this study, the evolution of the microstructure of the AlSi10Mg processed by LPBF, under a DSC heating thermal cycle,
was computed by an extended Phase Field KKS model. Thanks to CALPHAD data, it offers a sophisticated tool
to simulate the microstructural changes with an acceptable fidelity. Compared to the few previous implemented KKS models applied on AlSi10Mg processed by LPBF, the model considers non-isothermal conditions with a mixed-mode growth regime for Si precipitates and takes into account Si solute diffusion enhanced by the quenched-in vacancies for a more physical description of microstructure evolution in LPBF materials.	
\par  The extended Phase Field KKS model further opens the following opportunities :
	\begin{itemize}
		\item to feed Phase-Field computed properties into a finite element model studying rapid solidification under LBPF processes. In this case, multi-scale simulations could enhance the process model accuracy.
	\item to guide the optimization of heat treatments either in-situ in the LPBF process or as post-treatment. 
	\item to study the interest to preheat the substrate during the LPBF process by treating nucleation through the integration of the classical nucleation theory relationships within the developed extended Phase Field KKS model.
\end{itemize}
\section*{Declaration of competing interest }
The authors declare that they have no known competing financial interests or personal relationships that could
have appeared to influence the work reported in this paper.
\section*{Data availability }
All the material and numerical data are available within the article: a home-made Open-Source code for phase-field simulations which is implemented in Python and incorporates optimized techniques for enhanced computational efficiency. Additionally, it includes a sophisticated post-processing toolkit, also developed in Python. The github repository for this project could be found here: \href{here}{https://github.com/SFETNI/Extended-KKS-Phase-Field-model.git}

\section*{Acknowledgments}
As Research Director of F.R.S-FNRS, A.M. Habraken acknowledges the support of this institution. CAREM of the University of Liege is acknowledged for providing SEM/EDS facilities. \\
The ULiège research council of Sciences and Techniques and Engineering research council are acknowledged for the post-doc IN IPD-STEMA 2019 grant and faculty post doc grant 2021 of Seifallah Fetni.\\
The authors wish to acknowledge the financial support of the European Fund for Regional Development and the Walloon Region under convention FEDER “Iawatha” and of the Walinnov Longlife AM project, convention n°1810016, funded by the Service Public de Wallonie - Economie Emploi Recherche (SPW-EER). The authors are also grateful to Mr O. Rigo (Sirris) and Dr S. Michotte (AnyShape), for their help with the fabrication of the samples.


\printbibliography

@article{KKS1999,
  author             = {Kim, S. G. and Kim, W. T. and Suzuki, T.},
  title              = {Phase-field model for binary alloys.},
  issue              = {6 Pt B},
  pages              = {7186-97},
  volume             = {60},
  print-issn         = {1063-651X},
  doi                = {https://doi.org/10.1103/physreve.60.7186},
  history            = {2002/04/24 10:00 [entrez]},
  journal            = {Physical review. E, Statistical physics, plasmas, fluids, and related interdisciplinary topics},
  month              = {Dec},
  title-abbreviation = {Phys Rev E Stat Phys Plasmas Fluids Relat Interdiscip Topics},
  year               = {1999},
}

@article{YANG2023111,
title = {A novel computational model for isotropic interfacial energies in multicomponent alloys and its coupling with phase-field model with finite interface dissipation},
journal = {Journal of Materials Science and Technology},
volume = {133},
pages = {111-122},
year = {2023},
issn = {1005-0302},
doi = {https://doi.org/10.1016/j.jmst.2022.04.057},
author = {Shenglan Yang and Jing Zhong and Jiong Wang and Jianbao Gao and Qian Li and Lijun Zhang},

}

@article{YANG202278,
title = {Clustering and precipitation in Al-Mg-Si alloys during linear heating},
journal = {Journal of Materials Science and Technology},
volume = {120},
pages = {78-88},
year = {2022},
issn = {1005-0302},
doi = {https://doi.org/10.1016/j.jmst.2021.11.062},
author = {Zi Yang and Igor Erdle and Chunhui Liu and John Banhart},
}

@book{Höhne_Hemminger_Flammersheim_2003,
 title={Differential Scanning Calorimetry},
 DOI={10.1007/978-3-662-06710-9},
 publisher={Springer Berlin Heidelberg},
 author={Höhne, G. W. H. and Hemminger, W. F. and Flammersheim, H.-J.},
 year={2003}
 }

@article{SCHOLLER2022114965,
title = {Phase-field modeling of crack propagation in heterogeneous materials with multiple crack order parameters},
journal = {Computer Methods in Applied Mechanics and Engineering},
volume = {395},
pages = {114965},
year = {2022},
issn = {0045-7825},
doi = {https://doi.org/10.1016/j.cma.2022.114965},

author = {Lukas Schöller and Daniel Schneider and Christoph Herrmann and Andreas Prahs and Britta Nestler},

}

@article{ZHANG2020113310,
title = {Efficient and accurate numerical scheme for a magnetic-coupled phase-field-crystal model for ferromagnetic solid materials},
journal = {Computer Methods in Applied Mechanics and Engineering},
volume = {371},
pages = {113310},
year = {2020},
issn = {0045-7825},
doi = {https://doi.org/10.1016/j.cma.2020.113310},
author = {Jun Zhang and Xiaofeng Yang},
}

@article{BAO2022108215,
title = {The role of defects on tensile deformation and fracture mechanisms of AM {A}l{S}i10{M}g alloy at room temperature and 250 °C},
journal = {Engineering Fracture Mechanics},
volume = {261},
pages = {108215},
year = {2022},
issn = {0013-7944},
doi = {https://doi.org/10.1016/j.engfracmech.2021.108215},
author = {Jianguang Bao and Zhengkai Wu and Shengchuan Wu and Dianyin Hu and Wei Sun and Rongqiao Wang},
}

@article{Hegde2008,
title = {Modification of eutectic silicon in {A}l–{S}i alloys},
journal = {Journal of Materials Science},
volume = {43},
pages = { pages3009–3027 },
year = {2008},
issn = {},
doi = {https://doi.org/10.1007/s10853-008-2505-5},
author = {Sathyapal Hegde and K. Narayan Prabhu},
}

@article{SHANKAR20044447,
title = {Nucleation mechanism of the eutectic phases in aluminum–silicon hypoeutectic alloys},
journal = {Acta Materialia},
volume = {52},
number = {15},
pages = {4447-4460},
year = {2004},
issn = {1359-6454},
doi = {https://doi.org/10.1016/j.actamat.2004.05.045},
author = {Sumanth Shankar and Yancy W Riddle and Makhlouf M Makhlouf},
}

@article{Fujikawa1978,
author = {Fujikawa, Shin-ichiro and Hirano, Ken-ichi and Fukushima, Yoshiaki},
doi = {10.1007/BF02663412},
issn = {0360-2133},
journal = {Metallurgical Transactions A},
month = {dec},
number = {12},
pages = {1811--1815},
title = {Diffusion of silicon in aluminum},
volume = {9},
year = {1978}
}

@article{MANTINA20094102,
title = {First principles impurity diffusion coefficients},
journal = {Acta Materialia},
volume = {57},
number = {14},
pages = {4102-4108},
year = {2009},
issn = {1359-6454},
doi = {https://doi.org/10.1016/j.actamat.2009.05.006},
author = {M. Mantina and Y. Wang and L.Q. Chen and Z.K. Liu and C. Wolverton},
}

@article{ADR98,
title = "Definition of thermochemical and thermophysical properties to provide a database for the development of new light alloys : Thermochemical database for light metal alloys",
journal = "European Commission, Directorate-General for Research and Innovation, Publications Office,",
volume = "",
pages = "",
year = "1998",
author = "Ansara, I.and Rand, M. and Dinsdale, A.",
}

@article{Campbell2014,
  doi = {10.1186/2193-9772-3-12},
  year = {2014},
  publisher = {Springer Science and Business Media {LLC}},
  volume = {3},
  number = {1},
  pages = {158--180},
  author = {Carelyn E Campbell and Ursula R Kattner and Zi-Kui Liu},
  title = {The development of phase-based property data using the {CALPHAD} method and infrastructure needs},
  journal = {Integrating Materials and Manufacturing Innovation}
}

@article{GU2021110812,
title = {On the phase-field modeling of rapid solidification},
journal = {Computational Materials Science},
volume = {199},
pages = {110812},
year = {2021},
issn = {0927-0256},
doi = {https://doi.org/10.1016/j.commatsci.2021.110812},
author = {Yijia Gu and Xiaoming He and Daozhi Han},
}

@article{LINDROOS2022103139,
title = {Dislocation density in cellular rapid solidification using phase field modeling and crystal plasticity},
journal = {International Journal of Plasticity},
volume = {148},
pages = {103139},
author = {A. Falahati and E. Povoden-Karadeniz and P. Lang and P. Warczok and E. Kozeschnik},
year = {2022},
issn = {0749-6419},
doi = {https://doi.org/10.1016/j.ijplas.2021.103139},
}

@article{yang_deibler_bradley_stefan_carroll_2018, 
title={Microstructure evolution and thermal properties of an additively manufactured, solution treatable {A}l{S}i10{M}g part},
 volume={33}, 
DOI={10.1557/jmr.2018.405}, 
number={23}, journal={Journal of Materials Research},
 publisher={Cambridge University Press}, 
author={Yang, Pin and Deibler, Lisa A. and Bradley, Donald R. and Stefan, Daniel K. and Carroll, Jay D.}, year={2018}, pages={4040–4052}}

@article{BOTTGER20096784,
title = {Phase-field simulation of microstructure formation in technical castings – A self-consistent homoenthalpic approach to the micro–macro problem},
journal = {Journal of Computational Physics},
volume = {228},
number = {18},
pages = {6784-6795},
year = {2009},
issn = {0021-9991},
doi = {https://doi.org/10.1016/j.jcp.2009.06.028},
author = {B. Böttger and J. Eiken and M. Apel},
}

@article{FISCHER2015164,
title = {Relaxation of a precipitate misfit stress state by creep in the matrix},
author = {F.D. Fischer and J. Svoboda and T. Antretter and E. Kozeschnik},
journal = {International Journal of Plasticity},
volume = {64},
pages = {164-176},
year = {2015},
issn = {0749-6419},
doi = {https://doi.org/10.1016/j.ijplas.2014.08.014},
}

@article{BOYD19711101,
title = {A calorimetric determination of precipitate interfacial energies in two {A}l-{C}u alloys},
journal = {Acta Metallurgica},
volume = {19},
number = {10},
pages = {1101-1109},
year = {1971},
issn = {0001-6160},
doi = {https://doi.org/10.1016/0001-6160(71)90042-3},
author = {J.D. Boyd and R.B. Nicholson},
}

@article{LI201574,
  doi = {10.1016/j.actamat.2015.05.017},
  year = {2015},
  month = aug,
  publisher = {Elsevier {BV}},
  volume = {95},
  pages = {74--82},
  author = {X.P. Li and X.J. Wang and M. Saunders and A. Suvorova and L.C. Zhang and Y.J. Liu and M.H. Fang and Z.H. Huang and T.B. Sercombe},
  title = {A selective laser melting and solution heat treatment refined Al{\textendash}12Si alloy with a controllable ultrafine eutectic microstructure and 25{\%} tensile ductility},
  journal = {Acta Materialia}
}

@article{KIMURA20161294,
  title ={ Microstructures and mechanical properties of A356 (AlSi7Mg0.3) aluminum alloy fabricated by selective laser melting},
  journal = {Materials and Design},
  volume = {89},
  pages = {1294-1301},
  year = {2016},
  issn = {0264-1275},
  doi = {https://doi.org/10.1016/j.matdes.2015.10.065},
  author = {Takahiro Kimura and Takayuki Nakamoto},
}

@article{LIAO20071121,
title = {Refinement of eutectic grains by combined addition of strontium and boron in near-eutectic {A}l–{S}i alloys},
journal = {Scripta Materialia},
volume = {57},
number = {12},
pages = {1121-1124},
year = {2007},
issn = {1359-6462},
author = {Hengcheng Liao and Min Zhang and Qichang Wu and Huipin Wang and Guoxiong Sun},
}

@article{KELLER2017244,
title = {Application of finite element, phase-field, and CALPHAD-based methods to additive manufacturing of {N}i-based superalloys},
journal = {Acta Materialia},
volume = {139},
pages = {244-253},
year = {2017},
issn = {1359-6454},
doi = {https://doi.org/10.1016/j.actamat.2017.05.003},
author = {Trevor Keller and Greta Lindwall and Supriyo Ghosh and Li Ma and Brandon M. Lane and Fan Zhang and Ursula R. Kattner and Eric A. Lass and Jarred C. Heigel and Yaakov Idell and Maureen E. Williams and Andrew J. Allen and Jonathan E. Guyer and Lyle E. Levine},

}

@article{KATZDEMYANETZ2020110505,
title = {High entropy Al0.5CrMoNbTa0.5 alloy: Additive manufacturing vs. casting vs. CALPHAD approval calculations},
journal = {Materials Characterization},
volume = {167},
pages = {110505},
year = {2020},
issn = {1044-5803},
doi = {https://doi.org/10.1016/j.matchar.2020.110505},
author = {A. Katz-Demyanetz and I.I. Gorbachev and E. Eshed and V.V. Popov and V.V. Popov and M. Bamberger},
}

@article{CHEN20135877,
title = {The effect of interrupted aging on the yield strength and uniform elongation of precipitation-hardened {A}l alloys},
journal = {Acta Materialia},
volume = {61},
number = {15},
pages = {5877-5894},
year = {2013},
issn = {1359-6454},
doi = {https://doi.org/10.1016/j.actamat.2013.06.036},
author = {Y. Chen and M. Weyland and C.R. Hutchinson},

}

@article{SANTOSMACIAS2020231,
title = {Influence on microstructure, strength and ductility of build platform temperature during laser powder bed fusion of {A}l{S}i10{M}g},
journal = {Acta Materialia},
volume = {201},
pages = {231-243},
year = {2020},
issn = {1359-6454},
doi = {https://doi.org/10.1016/j.actamat.2020.10.001},
author = {Juan Guillermo {Santos Macías} and Thierry Douillard and Lv Zhao and Eric Maire and Grzegorz Pyka and Aude Simar},
}

@book{Meh07,
  doi = {10.1007/978-3-540-71488-0},
  year = {2007},
  publisher = {Springer Berlin Heidelberg},
  author = {Helmut Mehrer},
  title = {Diffusion in Solids}
}

@article{DELAHAYE2019160,
title = {Influence of Si precipitates on fracture mechanisms of {A}l{S}i10{M}g parts processed by Selective Laser Melting},
journal = {Acta Materialia},
volume = {175},
pages = {160-170},
year = {2019},
issn = {1359-6454},
doi = {https://doi.org/10.1016/j.actamat.2019.06.013},
author = {J. Delahaye and J. Tchoufang Tchuindjang and J. Lecomte-Beckers and O. Rigo and A.M. Habraken and A. Mertens},
}

@article{HUANG2020101693,
title = {Modeling on the molar volume of the {A}l–{C}u–{M}g–{S}i system},
journal = {Calphad},
volume = {68},
pages = {101693},
year = {2020},
issn = {0364-5916},
doi = {https://doi.org/10.1016/j.calphad.2019.101693},
author = {Dandan Huang and Shuhong Liu and Yong Du},
}

@article{HU2007303,
title = {Thermodynamic description and growth kinetics of stoichiometric precipitates in the phase-field approach},
journal = {Calphad},
volume = {31},
number = {2},
pages = {303-312},
year = {2007},
issn = {0364-5916},
doi = {https://doi.org/10.1016/j.calphad.2006.08.005},
author = {S.Y. Hu and J. Murray and H. Weiland and Z.K. Liu and L.Q. Chen},
}

@article{LASKOWSKI2021158630,
title = {Phase field model for multiphase alloys under arbitrary thermal history: An application to IN718 super-alloy},
journal = {Journal of Alloys and Compounds},
volume = {861},
pages = {158630},
year = {2021},
issn = {0925-8388},
doi = {https://doi.org/10.1016/j.jallcom.2021.158630},
author = {Robert Laskowski and Kun Wang and Rajeev Ahluwalia and Kewu Bai and Guglielmo Vastola and Yong-Wei Zhang},

}

@Book{PE92,
  author    = {Porter, David A.; Easterling, K. E.},
  publisher = { Springer US},
  title     = {Phase Transformations in Metals and Alloys},
  year      = {1992},
  isbn      = {978-0-442-31638-9 978-1-4899-3051-4},
}

@inproceedings{Su201507,
  title={Establishment of the Elastic Property Database of {F}e-base Alloys},
  author={Di Su and Yan-Lin He and Ji-Qiong Liu and Xiao-Gang Lu},
  year={2015},
  booktitle={Proceedings of the First International Conference on Information Sciences, Machinery, Materials and Energy},
  pages={1825-1835},
  issn={1951-6851},
  isbn={978-94-62520-67-7},
  doi={https://doi.org/10.2991/icismme-15.2015.377},
  publisher={Atlantis Press}
}

@article{HALLSTEDT2007292,
title = {Molar volumes of Al, Li, Mg and Si},
journal = {Calphad},
volume = {31},
number = {2},
pages = {292-302},
year = {2007},
issn = {0364-5916},
doi = {https://doi.org/10.1016/j.calphad.2006.10.006},
author = {Bengt Hallstedt},
}

@article{Falahati2010,
title = {Thermo-kinetic computer simulation of differential scanning calorimetry curves of {A}l{M}g{S}i alloys},
author = {A. Falahati and E. Povoden-Karadeniz and P. Lang and P. Warczok and E. Kozeschnik},
pages = {1089--1096},
volume = {101},
number = {9},
journal = {International Journal of Materials Research},
doi = {doi:10.3139/146.110396},
year = {2010},
}

@book{Hes10,
  title = {The Essential Guide to Learning Blender 2.6},
  author = {R. Hess},
  year = {2010},
  pages = {1},
  isbn = {978-0-240-81430-8},
  publisher = {Blender Foundations}
}

@phdthesis{hu2004phase,
  title={Phase-field Models of Microstructure Evolution in a System with Elastic Inhomogeneity and Defects},
  author={Hu, Shenyang},
  school={The Pennsylvania State University},
  year={2004},
  type={Ph.D. thesis},
  department={Materials Science and Engineering},
  note={Submitted in Partial Fulfillment of the Requirements for the Degree of Ph.D. in Materials Science and Engineering}
}

@article{FETNI2023111820,
title = {Capabilities of Auto-encoders and Principal Component Analysis of the reduction of microstructural images; Application on the acceleration of Phase-Field simulations},
journal = {Computational Materials Science},
volume = {216},
pages = {111820},
year = {2023},
issn = {0927-0256},
doi = {https://doi.org/10.1016/j.commatsci.2022.111820},
author = {Seifallah Fetni and Thinh Quy Duc Pham and Truong Vinh Hoang and Hoang Son Tran and Laurent Duchêne and Xuan-Van Tran and Anne Marie Habraken},
}

@article{Zhang2016,
title = {Thermal conductivity of Al–Cu–Mg–Si alloys: Experimental measurement and CALPHAD modeling},
journal = {Thermochimica Acta},
volume = {635},
pages = {8-16},
year = {2016},
issn = {0040-6031},
doi = {https://doi.org/10.1016/j.tca.2016.04.019},
url = {https://www.sciencedirect.com/science/article/pii/S0040603116300892},
author = {Cong Zhang and Yong Du and Shuhong Liu and Yuling Liu and Bo. Sundman},
}

@article{FETNI2021109661,
title = {Thermal model for the directed energy deposition of composite coatings of 316L stainless steel enriched with tungsten carbides},
journal = {Materials and Design},
volume = {204},
pages = {109661},
year = {2021},
issn = {0264-1275},
doi = {https://doi.org/10.1016/j.matdes.2021.109661},
author = {Seifallah Fetni and Tommaso Maurizi Enrici and Tobia Niccolini and Hoang Son Tran and Olivier Dedry and Laurent Duchêne and Anne Mertens and Anne Marie Habraken},
}

@inproceedings{Fetni2021_COMPLAS,
  doi = {10.23967/complas.2021.009},
  year = {2021},
  publisher = {{CIMNE}},
  author = {S. Fetni and J. Delahaye and L. Duch{\^{e}}ne and A. Mertens and A. Habraken},
  title = {Adaptive time stepping approach for Phase-Field modeling of phase separation and precipitates coarsening in additive manufacturing alloys},
  booktitle = {16th edition of the International Conference on Computational Plasticity}
}

@PhdThesis{Hu2004,
  author = {Hu, Shenyang},
  school = {The Pennsylvania State University},
  title  = {Phase-field Models of Microstructure Evolution in a System with Elastic Inhomogeneity and Defects},
  year   = {2004},
}

@article{GUILLENGONZALEZ2014821,
title = {Second order schemes and time-step adaptivity for {A}llen–{C}ahn and Cahn–Hilliard models},
journal = {Computers and Mathematics with Applications},
volume = {68},
number = {8},
pages = {821-846},
year = {2014},
issn = {0898-1221},
doi = {https://doi.org/10.1016/j.camwa.2014.07.014},
author = {Francisco Guillén-González and Giordano Tierra},
}

@article{MOULINEC199869,
title = {A numerical method for computing the overall response of nonlinear composites with complex microstructure},
journal = {Computer Methods in Applied Mechanics and Engineering},
volume = {157},
number = {1},
pages = {69-94},
year = {1998},
issn = {0045-7825},
doi = {https://doi.org/10.1016/S0045-7825(97)00218-1},
author = {H. Moulinec and P. Suquet},
}

@Book{Bin17,
  author    = {S. Bulent Biner},
  publisher = {Springer, Cham},
	doi = {https://doi.org/10.1007/978-3-319-41196-5},
  title     = {Programming Phase-Field Modeling},
  year      = {2017},
}

@article{AAGESEN201710,
title = {Quantifying elastic energy effects on interfacial energy in the {K}im-{K}im-{S}uzuki phase-field model with different interpolation schemes},
journal = {Computational Materials Science},
volume = {140},
pages = {10-21},
year = {2017},
issn = {0927-0256},
author = {Larry K. Aagesen and Daniel Schwen and Karim Ahmed and Michael R. Tonks},
}

@article{HAGHANIHASSANABADI2021110111,
title = {Phase-change modeling based on a novel conservative phase-field method},
journal = {Journal of Computational Physics},
volume = {432},
pages = {110111},
year = {2021},
issn = {0021-9991},
doi = {https://doi.org/10.1016/j.jcp.2021.110111},
author = {Reza Haghani-Hassan-Abadi and Abbas Fakhari and Mohammad-Hassan Rahimian},
}

@article{BOISSE20076151,
title = {Phase-field simulation of coarsening of $\gamma$ precipitates in an ordered gamma' matrix},
journal = {Acta Materialia},
volume = {55},
number = {18},
pages = {6151-6158},
year = {2007},
issn = {1359-6454},
doi = {https://doi.org/10.1016/j.actamat.2007.07.014},
author = {J. Boisse and N. Lecoq and R. Patte and H. Zapolsky},
}

@article{JI201884,
title = {Phase-field modeling of $\theta$'precipitation kinetics in 319 aluminum alloys},
journal = {Computational Materials Science},
volume = {151},
pages = {84--94},
year = {2018},
issn = {0927-0256},
doi = {https://doi.org/10.1016/j.commatsci.2018.04.051},
author = {Yanzhou Ji and Bita Ghaffari and Mei Li and Long-Qing Chen},
}



\section*{Supplementary material}

# Supplementary Material to the article: Extension of a phase-field KKS model to predict the microstructure evolution in LPBF AlSi10Mg alloy submitted to non isothermal processes

June 23, 2024

# Appendices

## A List of symbols

| Symbol | Designation | Unit |
|---|---|---|
| **Roman** | | |
| $a^\phi$ | Lattice parameter of the phase $\phi$ | [m] |
| $B^\alpha$ | Bulk modulus | [Pa] |
| $C_{ijkl}$ | Stiffness tensor | [Pa] |
| $C_p$ | Heat capacity | $[J.K^{-1}]$ |
| $C_{11}^\alpha, C_{12}^\alpha, C_{44}^\alpha$ | Elastic constants of the $\alpha$ phase | [Pa] |
| $\widetilde{D}$ | Interdiffusivity | $[m^2 s^{-1}]$ |
| $D_{Al}^{*\,\alpha}$ | Al self-diffusion coefficient | $[m^2 s^{-1}]$ |
| $D_i^{*\,\alpha}$ | Tracer or impurity diffusion coefficient of element i in $\alpha$ phase | $[m^2 s^{-1}]$ |
| $D_{Va}$ | Vacancy diffusion coefficient | $[m^2 s^{-1}]$ |
| $e_{el}$ | Elastic energy | $[J.m^{-3}]$ |
| $E^\alpha$ | Temperature dependent Young modulus of the $\alpha$-Al single crystal | $[Pa]$ |
| $f^\alpha$ | Molar Gibbs free energy of the $\alpha$-Al phase | $[J.mol^{-1}]$ |
| $f^\theta$ | Molar Gibbs free energy of the d-Si phase | $[J.mol^{-1}]$ |
| $f_0$ | Homogeneous Helmholtz free energy density | $[J.m^{-3}]$ |
| $f_2$ | Impurity correlation factor | [–] |
| $\mathcal{F}_{KKS}$ | Total Helmholtz free energy | $[J]$ |
| $g$ | Double well potential function | [–] |
| ${}^0 G_i^\phi$ | Molar Gibbs free energy of phase $\phi$ containing pure element i | $[J.\text{mol}^{-1}]$ |
| $G_m^\phi$ | Molar Gibbs free energy of phase $\phi$ | $[J.\text{mol}^{-1}]$ |
| $G_{mVa}^f$ | Molar Gibbs energy of vacancy formation | $[J.\text{mol}^{-1}]$ |
| $h$ | Interpolation function | [–] |

1

| | | |
|---|---|---|
| $H_m$ | Molar enthalpy of the system | [J.mol$^{-1}$] |
| $k$ | Quadratic norm of the Fourier space vector | [rad.m$^{-1}$] |
| $L$ | Interface kinetic coefficient | [J$^{-1}$s$^{-1}$] |
| $L_{Va}$ | Mean diffusion distance of the vacancies to the sources and sinks | [m] |
| $M$ | Mobility coefficient | [m$^2$.J$^{-1}$.s$^{-1}$] |
| $M_{Si}^{\alpha}$ | Atomic mobility of Si in the $\alpha$-Al matrix | [m$^2$.J$^{-1}$.s$^{-1}$] |
| $M_{Al}^{\alpha}$ | Atomic mobility of Al in the $\alpha$-Al matrix | [m$^2$.J$^{-1}$.s$^{-1}$] |
| $M_\eta$ | Interface mobility | [m$^4$.J$^{-1}$.s$^{-1}$] |
| $N_p^\phi$ | Molar fraction of $\phi$ phase | [m$^{-2}$] |
| $Q_{D_{Si}^{*\alpha}}$ | Activation energy for the diffusion, $D_{Si}^{*\alpha}$[m$^2$.s$^{-1}$] is the frequency factor | [J.mol$^{-1}$] |
| $\dot{q}_W$ | Integral heat flux per unit of mass | [J.g$^{-1}$.s$^{-1}$] |
| $r_d$ | Equivalent radius of d-Si precipitates | [nm] |
| $RES_{KKS}$, resmin, resmax | Energy residuals | [J.s$^{-1}$] |
| $T$ | Temperature | [K] |
| $t$ | Time | [s] |
| $V$ | Volume of the system | [m$^3$] |
| $V_d$ | Volume of the d-Si phase | [m$^3$] |
| $V^{d_0}$ | Initial volume fraction of d-Si precipitates | [–] |
| $V_{m_0}^\phi$ | Molar volume of $\phi$ phase at 298 K | [m$^3$.mol$^{-1}$] |
| $\langle V_m \rangle$ | Mean molar volume of the system | [m$^3$.mol$^{-1}$] |
| $w$ | Double-well height | [J.mol$^{-1}$] |
| $w_2$ | Vacancy impurity exchange rate | [s$^{-1}$] |
| $X_{Si}^\alpha$ | Molar fraction of Si in $\alpha$-phase | [–] |
| $X_{Si}^d$ | Molar fraction of Si in d-Si phase | [–] |
| $X_i^{\phi_0}$ | Initial molar fraction of an element i in the phase $\phi$ | [–] |
| $X_i^{\phi_e}$ | Molar fraction of an element i in the phase $\phi$ at equilibrium | [–] |
| $X_i^{\alpha_i}$ | Si molar fraction profile at the d-Si precipitate interface | [–] |
| $X_{Si}$ | Molar fraction of Si in the system | [–] |
| $X_{Va}$ | Vacancy site fraction | [–] |
| $X_{Va}^e$ | Vacancy site fraction at equilibrium | [–] |
| $\langle X_{Si}^\alpha \rangle$ | Mean volume fraction of Si precipitates in $\alpha$ phase | [–] |
| $z_L$ | Number of nearest neighbors in the precipitate | [–] |
| $z_S$ | Number of nearest neighbors at the surface of the precipitate | [–] |
| $\{.\}_k$ | Fourier transform operator | [–] |
| $[.]_r$ | Inverse Fourier transform operator | [–] |

**Greek**

| | | |
|---|---|---|
| $\alpha_1$ | Thermal diffusivity | [m$^2$.s$^{-1}$] |
| $\alpha$ | A constant determined by the interface definition | [–] |
| $\chi$ | Proportionality parameter in mixed-mode analytical model | [J.mol$^{-1}$] |
| $\Delta t$ | Time step | [s] |
| $\Delta x - \Delta y$ | Grid spacing | [m] |
| $\delta$ | Kronecker's symbol | [–] |
| $\varepsilon_{ij}$ | Total strain tensor | [–] |
| $\varepsilon_{ij}^{el}$ | Elastic strain tensor | [–] |
| $\varepsilon_{ij}^{0}$ | Stress-free strain tensor | [–] |

| | | |
|---|---|---|
| $\phi_c$ | Thermodynamic function | $[J]$ |
| $\gamma$ | Interfacial energy | $[J.m^{-2}]$ |
| $\gamma_H$ | Interfacial enthalpy | $[J.m^{-2}]$ |
| $\gamma_S$ | Interfacial entropy | $[J.m^{-2}.K^{-1}]$ |
| $\eta$ | Order parameter (d-Si diamond phase) | $[-]$ |
| $\langle\eta\rangle$ | Mean d-Si phase fraction | $[-]$ |
| $\kappa$ | Gradient energy coefficient | $[J^{1/2}.m^{-3/2}]$ |
| $2\lambda$ | Interface thickness | $[m]$ |
| $\lambda_1$ | Thermal conductivity | $[W.m^{-1}.K^{-1}]$ |
| $\mu_i^\phi$ | Chemical potential of element i in phase $\phi$ | $[J.mol^{-1}]$ |
| $\mu_m^\phi$ | Chemical molar potential of the phase $\phi$ | $[J.mol^{-1}]$ |
| $\theta$ | Coefficient for time step correction | $[-]$ |
| $\rho$ | Density | $[Kg.m^{-3}]$ |
| $\sigma_{ij}$ | Elastic tress tensor | $[Pa]$ |
| $\sigma_{ij}^0$ | Eigen stress tensor | $[Pa]$ |
| $^\nu\Omega_{ij}^\phi$ | Gibbs free energy binary interaction parameter | $[J.mol^{-1}]$ |
| $*$ | Dimensionless gradient operator | $[-]$ |

# B  Model parameters and CALPHAD data

Some mathematical developments of formulas of section 5, as well as CALPHAD data and details about numerical implementation, are presented hereafter.

## B.1  Gibbs free energy curves

Gibbs free energy curves for $\alpha-Al$ ($G_m^\alpha$) and d-Si ($G_m^d$) phases as well as their parabola approximations ($^PG_m^\alpha, ^PG_m^d$) are presented in Fig. B1.1 for $T$=850 K. The values are computed with pure $\alpha$-Al and $d$-Si phase as reference states for the given temperature. One can see that the Gibbs free energy parabola of $\alpha$-Al phase approximates the CALPHAD one for the Si molar fraction range of interest i.e. $X_{Si}^{\alpha_e} \leqslant X_{Si}^\alpha \leqslant X_{Si}^{\alpha_0}$ with $X_{Si}^{\alpha_0} = 0.025$. Regarding the d-Si phase, since the coefficient $^PA^d$ does not influence the transformation kinetics in the phase-field model ([1]), $^PA^d = {}^PA^\alpha$ is enforced and the simplification in Eq. 17 is used. The coefficients of the parabola approximation of the $\alpha$-Al and d-Si phase molar Gibbs free energy can be found in Table B1.1.

| $T$ | $^PA^\alpha$ | $^PB^\alpha$ | $^PC^\alpha$ | $^PA^d$ | $^PB^d$ | $^PC^d$ |
|---|---|---|---|---|---|---|
| $[K]$ | | | | $[J.mol^{-1}]$ | | |
| 400 | 979342 | $-18.9$ | $-0.023$ | 979342 | $-1958665$ | 979323 |
| 490 | 759537 | $-181.3$ | $-0.408$ | 759537 | $-1519059$ | 759522 |
| 580 | 562013 | $-805.7$ | $-2.788$ | 562013 | $-1124012$ | 561999 |
| 670 | 405520 | $-2184.1$ | $-10.525$ | 405520 | $-811019$ | 405499 |
| 760 | 298132 | $-4381.6$ | $-25.997$ | 298132 | $-596215$ | 298083 |
| 850 | 247099 | $-7964.1$ | $-40.079$ | 247099 | $-494088$ | 246988 |

Table B1.1: Parabola coefficients of the molar Gibbs free energy.

3

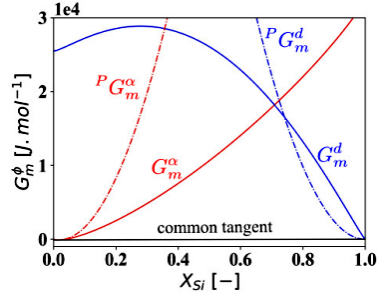

Figure B1.1: Free energy curves of $\alpha$-Al and d-Si phases at a temperature $T$=850 K. The solid lines represent data from CALPHAD whereas dotted lines represent parabola approximations.

## B.2 Interfacial energy

For the determination of the interfacial energy, a thermodynamic approach could be chosen such as the nearest neighbour broken bond (NNBB) model proposed in [2]:

$$\gamma = \frac{n_S\,(z_L - z_S)}{N_a z_L}\frac{\partial H_m}{\partial N_p^d} \tag{B2.1}$$

where $n_S$ is the number of atoms per unit surface of one precipitate, $z_L$ and $z_S$ are the numbers of nearest atom neighbours in the precipitate and at the surface of the precipitate respectively. For a the face-centered d-Si lattice in 3D, $z_L$=12 and $z_S$ is taken as an average of crystallographic planes $[100], [110]\,and\,[111]$, $(z_S^{[100]} + z_S^{[110]} + z_S^{[111]})/3$= (8+7+9)/3=8.
The number of atoms per unit surface of precipitate $V_m^d$ is calculated from the molar volume of the precipitate as follows:

$$n_S = \left(\frac{V_m^d}{N_a}\right)^{-2/3} \tag{B2.2}$$

The molar enthalpy of the system is derived from the Gibbs free energy from [3]:

$$H_m = G_m - T\left(\frac{\partial G_m}{\partial T}\right)_{P,X_i} \tag{B2.3}$$

The molar enthalpy of the system could be also expressed from the molar enthalpy $H_m^\phi$ of the phases that constitute the system:

$$H_m = \sum_\phi N_p^\phi H_m^\phi \tag{B2.4}$$

where $N_p^\phi$ is the molar fraction of phase $\phi$.

## B.3 Enhanced diffusion by quenched-in excess vacancies

The vacancy site fraction computed from Eq. 26 with respect to different cooling rates is shown in Fig. B3.1. The curves were calculated with the concentration of Si in $\alpha$ phase $X_{Si}^\alpha$ at equilibrium given by the Al-Si phase diagram in (Eq. 25). The black curve represents the

4

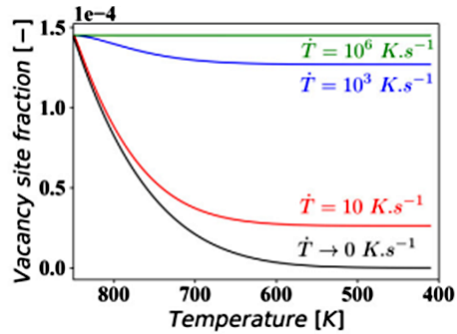

Figure B3.1: Vacancy site fraction for different values of cooling rate.

equilibrium vacancy site fraction. One can see that almost all the vacancies are entrapped for a cooling rate higher than $10^3$ K.$s^{-1}$. This cooling rate is the lower limit encountered during the LPBF process. Thus, the effect of the quenched-in excess vacancies has to be taken into account in the phase-field model. We should notice that at low temperature, only a tiny fraction of vacancies is present at equilibrium and thus the quantity $\frac{X_{Va}}{X_{Va}^e}$ becomes high. As seen in section 5, the time step in the phase-field model is normalized with the modified impurity diffusion coefficient incorporating the quenched-in vacancies from Eq. 25. In that case, the time step can be very small. In order to speed up the computation, a threshold value $tr_{Va}$ is enforced to the quantity $\frac{X_{Va}}{X_{Va}^e}$ such as:

$$\frac{X_{Va}}{X_{Va}^e} = \begin{cases} \frac{X_{Va}}{X_{Va}^e} & \text{if } \frac{X_{Va}}{X_{Va}^e} < tr_{Va} \\ tr_{Va} & \text{if } \frac{X_{Va}}{X_{Va}^e} > tr_{Va} \end{cases} \tag{B3.1}$$

## C Sensitivity analysis

The sensitivity of the results of the extended KKS model to the material and numerical parameters was performed based on a reference simulation. The values of the spatio-temporal parameters as well as the material parameter coefficients of this reference simulation are shown in Table C.1. The initial reference microstructure and its evolution under heating are presented in Fig.C.1 a and b respectively. It should be noted that the choice of a smaller grid than the one of section 6.1 allows faster simulations.

| Calibration coefficients for material parameters | | | | | | | | | |
|---|---|---|---|---|---|---|---|---|---|
| Interface mobility coefficient $f_{M_\eta}$ | | Diffusivity coefficient $f_{\widetilde{D}}$ | | Interfacial energy coefficient $f_\gamma$ | | Elastic energy coefficient $f_{\varepsilon_0}$ | | Kinetic coefficient of the interface $f_L$ | |
| Ref | Range | Ref | Range | Ref | Range | Ref | Range | Ref | Range |
| 1 | 0.5 − 1.5 | 1 | 0.5 − 1.5 | 2.15 | 1.15 − 3.15 | 0.01 | 0.1 − 0.001 | 0.01 | 0.001 − 0.1 |

Table C.1: Parameters studied in the sensitivity analysis with their variation range.

The sensitivity analysis consisted in varying the values of the material parameters, to understand the sensitivity of the results to their variation as detailed in Table C.1.

5

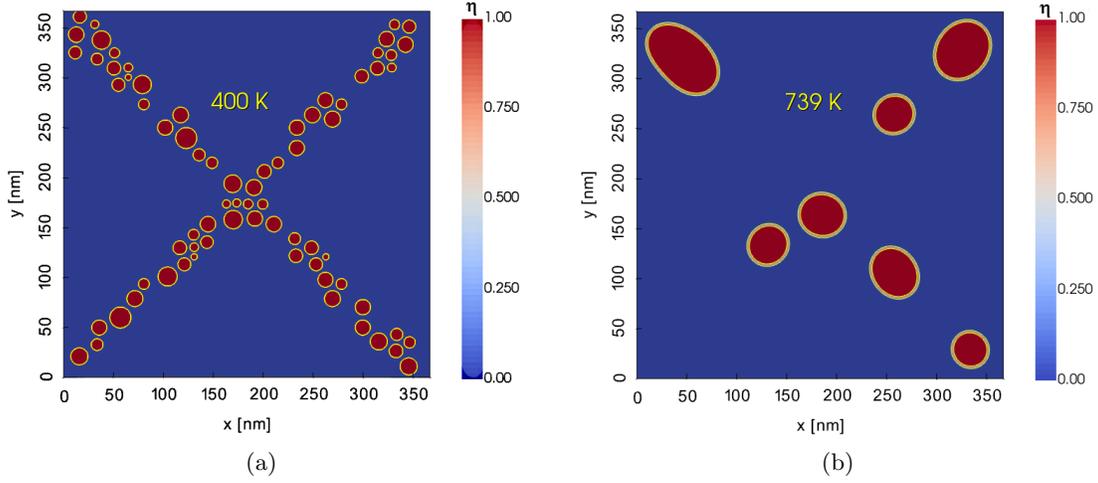

Figure C.1: Reference 2D maps of the d-Si phase fraction/order parameter ($\eta$) at the initial state and 400 K (a) and at 739 K (b) for a simulated DSC heating of 20K/min.

The results are shown in Fig.C.2. It should be inferred that no noticeable changes are observed when varying the values of $f_{\tilde{D}}$ and $f_{\varepsilon 0}$ around the reference solution for the first DSC peak due to the Si precipitation. However, the second peak position, due to the precipitate coalescence, is shifted to high temperatures. For the coefficients $f_\gamma$ and $f_{M\eta}$, not only there is a shift in the second peak position, but also noticeable changes specially in their amplitude and shapes can be observed. The most influential material parameter coefficient is the kinetic coefficient of the interface $f_L$, because there is not only a change in the shape of the peaks, but also in their position. These observations confirm the strong impact of the diffusivity coefficient ($f_{\tilde{D}}$) and the coefficient of interfacial elastic energy coefficient ($f_\gamma$) on the specific heat for the Si precipitation. On the other hand, the interface mobility coefficient ($f_{M\eta}$) and kinetic coefficient of the interface ($f_L$,) have a strong impact on the precipitate coalescence, since depending on these values the temperature required for coalescence significantly changes and for wrong value do not represent the experimental observations.

The sensitivity of the computed results to the adaptive time stepping scheme should also be highlighted. Indeed, once the initial time step and its multiplicative factor or divisor $\theta$ were defined (see Table 2) allowing the converged simulations to predict consistent physical results, the set of values of the adaptive scheme (resmin = $10^{-4}$, resmax= $10^{-1}$, and $iter_{max}$ =15) was carefully chosen in order to ensure fast simulations and guarantee the stability of the computed results. With the optimal set of parameters identified, the scheme is able to provide fast and stable simulations and requires a minimum number of iterations (up to 7 iterations) to compute the correct equilibrium solution which is granted by a low value of $RES_{KKS}$ value (Eq. 36). The values of resmin and resmax are significantly improved compared to those presented in [4]. Meanwhile, as reported in [4], the adaptive time stepping scheme values strongly depend on the material and discretization parameters (time step, grid spacing, resmin and resmax).

6

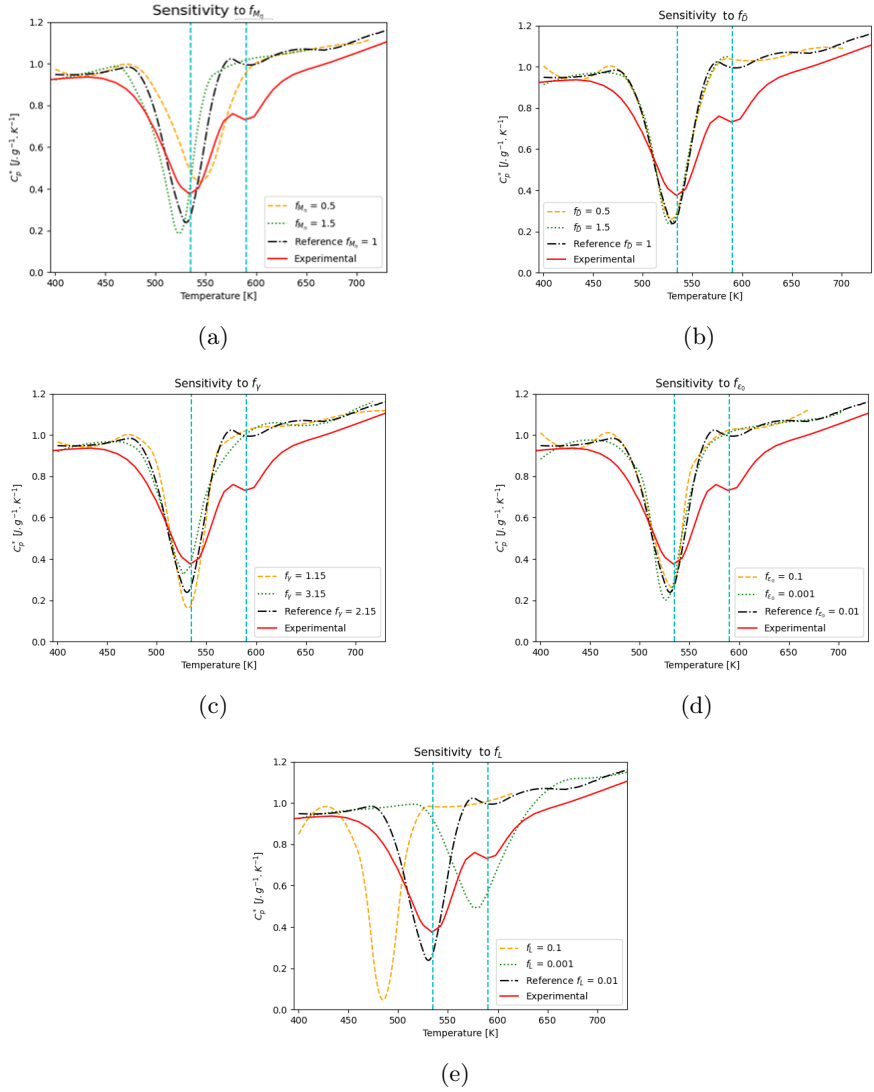

Figure C.2: DSC curves with different values of materials parameters: (a) interface mobility $f_{M\eta}$, (b) diffusivity coefficient $f_{\tilde{D}}$, (c) interfacial energy coefficient $f_\gamma$, (d) elastic energy coefficient $f_{\varepsilon 0}$, and (e) kinetic coefficient of the interface $f_L$

# D  Animation of the micro-structure Evolution

The following is the supplementary data related to this article. The above clip illustrates how the microstructure of the LPBF AlSi10Mg alloy evolves during a simulated DSC test.

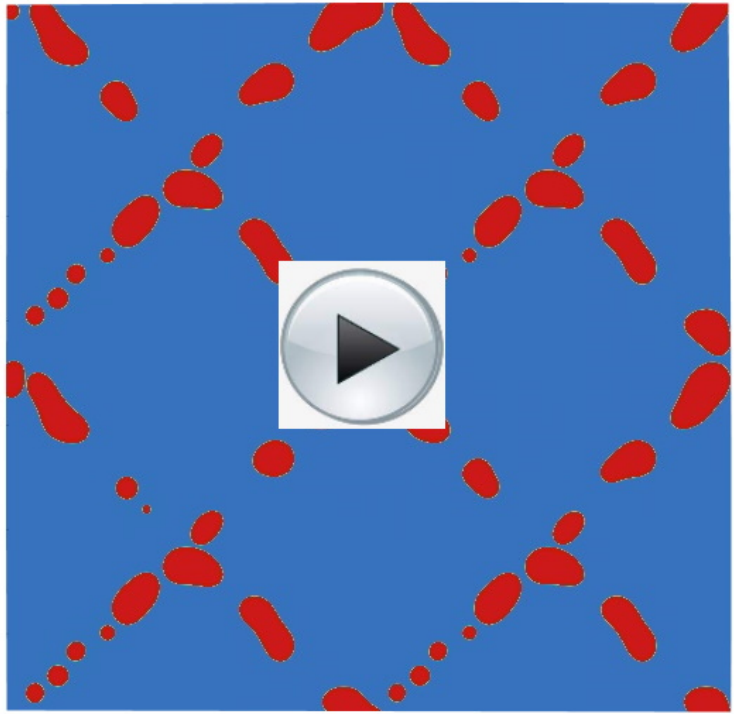

Figure D.1: Microstructure evolution during a simulated DSC test.

# References

[1] S. Hu, J. Murray, H. Weiland, Z. Liu, L. Chen, Thermodynamic description and growth kinetics of stoichiometric precipitates in the phase-field approach, Calphad 31 (2) (2007) 303–312. doi: https://doi.org/10.1016/j.calphad.2006.08.005.

[2] H. I. . S. Kozeschnik, E., On the potential for improving equilibrium thermodynamic databases with kinetic simulations, J Phs Eqil and Diff 28 (2007) 64–71. doi:https://doi.org/10.1007/s11669-006-9003-8.

[3] H. Lukas, S. G. Fries, B. Sundman, Computational Thermodynamics : The Calphad Method, Cambridge ; New York: Cambridge University Press, 2007. doi:https://doi.org/10.1017/CBO9780511804137.

[4] F. Guillén-González, G. Tierra, Second order schemes and time-step adaptivity for Allen–Cahn and cahn–hilliard models, Computers Mathematics with Applications 68 (8) (2014) 821–846. doi:https://doi.org/10.1016/j.camwa.2014.07.014.



\end{document}